# Multiple penalized least squares and sign constraints with modified Newton-Raphson algorithms: application to EEG source imaging.

**Short running title:** MNR algorithms for EEG source imaging.

Mayrim Vega-Hernández[1,2,a], José M. Sánchez-Bornot[3,a], Agustín Lage-Castellanos[2], Jhoanna Pérez-Hidalgo-Gato[2], Darío Palmero-Ledón[2], José E. Álvarez-Iglesias[2], Daysi García-Agustin[4], Eduardo Martínez-Montes[2,b], Pedro A. Valdés-Sosa[1,2,b].

[1] The Clinical Hospital of Chengdu Brain Science Institute, MOE Key Lab for Neuroinformation, University of Electronic Science and Technology of China, Chengdu, China.

[2] Cuban Center for Neurosciences, Havana, Cuba.

[3] School of Computing and Intelligent Systems, Ulster University, UK.

[4] Cuban Centre for Longevity, Ageing and Health Studies, Cuba.

[a] contributed equally to this research

[b] corresponding author.

**Abstract:**

Multiple penalized least squares (MPLS) models are a flexible approach to find adaptive least squares solutions required to be simultaneously sparse and smooth. This is particularly important when addressing real-life inverse problems where there is no ground truth available, such as electrophysiological source imaging. In this work we formalize a modified Newton-Raphson (MNR) algorithm to estimate general MPLS models and propose its extension to perform efficient optimization over the active set of selected features (AMNR). This algorithm can be used to minimize continuously differentiable objective functions with multiple restrictions, including sign constraints. We show that these algorithms provide solutions with acceptable reconstruction in simulated scenarios that do not cope with model assumptions, and for low n/p ratios. We then use both algorithms for estimating different electroencephalography (EEG) inverse models with multiple penalties. We also show how the AMNR allows us to estimate new models in the EEG inverse problem context, such as nonnegative versions of Smooth Garrote and Smooth LASSO. Synthetic data were used for a comparison between solutions obtained using the two algorithms proposed here and the least angle regression (LARS) algorithm, according to well-known quality measures. A visual event-related EEG from healthy young subjects and a resting-state EEG study on the relationship between cognitive ageing and walking speed decline in active elders, were used to illustrate the usefulness of the proposed methods in the analysis of real experimental data.

## 1. Introduction

Linear models are widely used due to their numerous applications. The linear regression model is stated as $\mathbf{y} = \mathbf{X}\boldsymbol{\beta} + \boldsymbol{\varepsilon}$, where the columns $(\mathbf{x}_1, \ldots, \mathbf{x}_p \in \mathbb{R}^n)$ of the design matrix $\mathbf{X}$ are the predictors, $\mathbf{y} \in \mathbb{R}^n$ is the response vector, $\boldsymbol{\beta} \in \mathbb{R}^p$ is the vector of coefficients to be estimated and $\boldsymbol{\varepsilon} \in \mathbb{R}^n$ is the error term, with the typical simple assumption that $\boldsymbol{\varepsilon} \sim \mathbf{N}(\mathbf{0}, \sigma^2 \mathbf{I}_n)$, where $\sigma^2$ is the variance of the noise component and $\mathbf{I}_n$ represents the size-$n$ identity matrix. When $p \gg n$ this model corresponds to an underdetermined system with no unique solution (ill-posed problem), which implies the need of introducing constraints to "select" a solution to the problem. This has led to a huge amount of scientific work on how to estimate models efficiently and reliably with different types of additional constraints, with many new extensions of regularization techniques in the last decade (Voronin (2012)).

These techniques produce biased but stable linear solutions when using L2 norm penalties, being Ridge regression (Hoerl and Kennard (1970)) the classical example. The advent of the least absolute shrinkage and selection operator (LASSO) (Tibshirani (1996)) and the emergence of the more general penalized least squares (PLS) formulation (Fan and Li (2001)) allowed the recovery of sparse solutions, where a large number of coefficients are forced to be zero by increased penalization, in contrast to Ridge regression which never produces sparse solutions. Methods producing sparse estimators are considered variable selection techniques in the PLS context. Moreover, the sparsity constraint can be naturally combined with other constraints to obtain estimators with simultaneous sparse, smooth, and possibly non-negative characteristics. In this broad sense, the Fused LASSO (Tibshirani et al. (2005)), the Fusion LASSO (FnLASSO) (Land and Friedman (1996)), the Elastic Net (ENET) (Zou and Hastie (2005)) and the Smooth LASSO (SLASSO) (Hebiri and van de Geer (2011)), can be seen as particular instances. A general model consisting of the combination of any number of penalty terms has been named as Multiple PLS (MPLS) (Vega-Hernández et al. (2008); Sánchez-Bornot et al. (2008)), and stated as the minimization of an objective function of the form:

$$f(\boldsymbol{\beta}) = (\mathbf{y} - \mathbf{X}\boldsymbol{\beta})^T(\mathbf{y} - \mathbf{X}\boldsymbol{\beta}) + \Psi(\boldsymbol{\beta}), \text{ with } \Psi(\boldsymbol{\beta}) = \sum_{r=1}^{R} \lambda_r G_r(\boldsymbol{\beta}) \qquad (1.1)$$

where the penalty term $\Psi(\boldsymbol{\beta})$ takes the form of a sum of $R$ convex constraints or penalty functions $G_r$, for $r = 1, \dots, R$. In turn, these functions can be written, in general, as a sum of functions on each component of a linear combination of the parameters $\boldsymbol{\beta}$, that is: $G_r(\boldsymbol{\beta}) = \sum_{i=1}^{N_r} g_r\left(\left|\theta_i^{(r)}\right|\right)$, where $\left|\theta_i^{(r)}\right|$ represents the absolute value of each component of the vector $\boldsymbol{\theta}^{(r)} = \mathbf{L}^{(r)}\boldsymbol{\beta}$, with $\mathbf{L}^{(r)} \in \mathbb{R}^{N_r \times p}$ being linear operators that impose a structural relationship among coefficients, (e.g. the matrix of first or second differences). In this work, we assume that $g_r: \mathbb{R} \mapsto \mathbb{R}$, are symmetric, non-negative, non-decreasing and continuous over $(0, +\infty)$, while the regularization parameters $\lambda_r$ establish the relative importance of each constraint. As can be easily shown, LASSO and Ridge regression are instances of equation (1.1), by setting $R = 1$, $\mathbf{L}^{(r)} = \mathbf{I}_p$ (the $p \times p$ identity matrix) and using the L1 and L2 norms as penalty functions, respectively. These and other particular examples are summarized in Table S1 of the Supplementary Material (online).

To obtain a reliable estimation of solutions in some particular instances of PLS models, many traditional approaches (e.g., conjugate-gradient, coordinate-wise descent and Newton-Raphson) have been used. Specifically, the Local Quadratic Approximation (LQA) (Fan and Li (2001)) and the Majorize-Minimize (MM) (Hunter and Li (2005); Lange (2016)) algorithms have provided a numerical engine to implement PLS methods. These algorithms can be seen as applications of a Newton-Raphson (NR) technique using an approximation of the objective function to produce sparse solutions, although using a numerical trick to enforce sparsity and to ensure numerical stability (Li et al. (2006)). However, to our knowledge, they haven't been formulated for a general MPLS model. Another algorithm for solving PLS models is Coordinate Descent (CD), implemented in the popular GLMNET package by Friedman (Friedman, Hastie and Tibshirani (2010)), which has been slightly improved by replacing each CD step with a Coordinate-wise Majorization Descent operation (Yang and Zou (2012)). Alternatively, the least angle regression (LARS) (Efron (2004)) and the Shooting algorithm (Fu (1998)) also known as coordinate-wise descent (Friedman et al. (2007)), offer efficient implementations for several of these PLS methods with the advantage that they make variable selection and estimation simultaneously. However, despite recent sophisticated algorithms (Rong et al. (2017)), their application scope is not as extensive as that of the LQA and MM approaches.

In addition to using several constraints for handling more complex real-world models, many applications require the use of nonnegative constraints on the solution. This is not an easy task and several attempts have been done since the introduction of the "best subset selection", which was one of the first variable selection procedures but cannot be represented as a PLS method (Hocking and Leslie (1967)). The instability of this method is well-known and thus, Breiman introduced the Nonnegative Garrote (NNG) as a variable selection technique that shrinks and zeroes the ordinary least squares (OLS) estimator, in order to give intermediate results between OLS and subset selection (Breiman (1995)), (see Table S1 of the Supplementary Material). Gijbels and colleagues introduced three robust versions of the nonnegative garrote, namely the M-, LTS, and S-nonnegative garrote (Gijbels and Vrinssen (2015)). They also introduced the MM-nonnegative garrote by combining the S- and M-nonnegative garrote. One of their findings is that the NNG can use other solutions as a reference solution instead of the OLS, leading to a different final estimator. However, one important limitation of NNG is that it is restricted to the $p < n$ case, which hinders its applications to general ill-posed inverse problems. Another approach was followed by Mørup (Mørup Madsen, and Hansen (2008)), who introduced a version of the LARS algorithm to implement the LASSO with nonnegative constraints. However, to our knowledge, algorithms for imposing nonnegativity has not been proposed within the general MPLS approach.

To deal with applications on real-world inverse problems where there is no experimental ground truth available (e.g., the EEG inverse problem), we have focused on the development of efficient algorithms to address flexible MPLS models which may produce smooth/sparse and/or sparse/sign-constrained solutions. In this work, we formalize a modified Newton-Raphson (MNR) algorithm, as a natural extension of the MM approach that allows solving MPLS models with similar convergence properties as those for MM and LQA algorithms. However, it does not offer nonnegative solutions, since the use of sign constraints within the MNR algorithms is not straightforward. Secondly, we propose an algorithm based on the active set technique for MPLS models, which will be called AMNR and can be seen as an extension of the LARS algorithm for convex and continuously differentiable cost functions in possible nonnegative (nonpositive) scenarios. We introduce the AMNR extension of the NNG to deal with the $p \gg n$ case, and to handle other novel combined methods that we call the Smooth Nonnegative Garrote (SNNG) and the Nonnegative Smooth LASSO (NN-SLASSO).

## 2. Modified Newton Raphson algorithm for MPLS models

In this section, we formalize a type of MNR algorithm to implement MPLS models. This method follows the philosophy of the MM algorithm introduced by Hunter and Lange (Hunter & Lange (2004)), who showed that the update formula for the MM algorithm is a particular case of the classical Newton-Raphson procedure. The MNR algorithm introduced in this article can be considered as an extension of the MM for dealing with MPLS models. Note that Hunter and Li (Hunter and Li (2005)) proposed a formulation for a penalized least squares (PLS) model that allows the use of different penalty weights for the components of $\boldsymbol{\beta}$. However, under their framework, it is not possible to combine several different penalty functions on the entire vector in the same penalized model.

The model raised by Hunter and Li (Hunter and Li (2005)) can be seen as a particular case of the general MPLS objective function in equation (1.1) when rewriting the penalty term as the sum over component-wise functions:

$$\Psi(\boldsymbol{\beta}) = \sum_{r=1}^{R} \lambda_r \sum_{i=1}^{N_r} g_r\left(\left|\theta_i^{(r)}\right|\right) = \sum_{i=1}^{N_r} \left(\sum_{r=1}^{R} \lambda_r g_r\left(\left|\theta_i^{(r)}\right|\right)\right)$$

Recall that we are using a general approach where constraints are applied to any linear combination of the parameters $\boldsymbol{\beta}$. Thus, the scalar magnitude $\theta_i^{(r)}$ is the $i$-th element of $\boldsymbol{\theta}^{(r)} = \mathbf{L}^{(r)}\boldsymbol{\beta}$ that models the row-wise correlation structure of $\boldsymbol{\beta}$ and generalizes the use of different weights for each of its components. Although the variables $\boldsymbol{\theta}^{(r)}$ are different in general, it is possible to derive a general algorithm using the local quadratic approximation proposed by Hunter and Li by perturbing every penalty function $g_r$. As shown in detail in Section 1 of the Supplementary Material (online), this allows us to use the Newton-Raphson technique to minimize the perturbed objective function $f_\varepsilon(\boldsymbol{\beta})$ through its first and second derivatives:

$$\nabla f_\varepsilon(\boldsymbol{\beta}) = -\mathbf{X}^T\mathbf{y} + \left(\mathbf{X}^T\mathbf{X} + \sum_{r=1}^R \lambda_r \mathbf{L}^{(r)T}\mathbf{D}^{(r)}\mathbf{L}^{(r)}\right)\boldsymbol{\beta}$$

$$\nabla^2 f_\varepsilon(\boldsymbol{\beta}) = \mathbf{X}^T\mathbf{X} + \sum_{r=1}^R \lambda_r \mathbf{L}^{(r)T}\mathbf{D}^{(r)}\mathbf{L}^{(r)}$$

where $\mathbf{D}^{(r)}$ is a diagonal matrix with diagonal elements defined as $d_i^{(r)} = g_r'\left(|\theta_i^{(r)}|\right)/\left(\varepsilon + |\theta_i^{(r)}|\right)$ for $i = 1, \ldots, N_r$ and some very small $\varepsilon > 0$. Then, we can locally minimize the perturbed objective function for some $\alpha_k > 0$, using the iterative formula:

$$\boldsymbol{\beta}_{k+1} = \boldsymbol{\beta}_k - \alpha_k\{\nabla^2 f_\varepsilon(\boldsymbol{\beta}_k)\}^{-1}\nabla f_\varepsilon(\boldsymbol{\beta}_k) = \boldsymbol{\beta}_k + \alpha_k\left[\left(\mathbf{X}^T\mathbf{X} + \sum_{r=1}^R \lambda_r \mathbf{L}^{(r)T}\mathbf{D}^{(r)}\mathbf{L}^{(r)}\right)^{-1}\mathbf{X}^T\mathbf{y} - \boldsymbol{\beta}_k\right]$$

Following the same rationale as Hunter and Li on the Majorize–Minimize algorithm we then obtain the canonical version of our MNR algorithm, which is shown in the table Algorithm 1. The basic difference with the MM is that, in step 5 of Algorithm 1, the sum of the derivatives of all penalty functions are included in the regularization of the inverse of the design matrix. This is precisely what makes this algorithm general for any differentiable penalty functions $g_r$ and linear operators $\mathbf{L}^{(r)}$.

---

**Algorithm 1. MNR for MPLS** $(\mathbf{y} \in \mathbb{R}^{n\times 1}, \mathbf{X} \in \mathbb{R}^{n\times p}, \lambda_1, \ldots, \lambda_R, \mathbf{L}^{(1)}, \ldots, \mathbf{L}^{(r)} \in \mathbb{R}^{N_r \times p})$

---

1. Start with $k: k \leftarrow 0$ and set $\tau \leftarrow 10^{-8}, \varepsilon \leftarrow 10^{-8}$, MaxIter$\leftarrow 100$ and $\boldsymbol{\Omega} \leftarrow \mathbf{I}_p$

2. Set $k \leftarrow k+1$ and compute $\boldsymbol{\beta}_k \leftarrow (\mathbf{X}^T\mathbf{X} + \boldsymbol{\Omega})^{-1}\mathbf{X}^T\mathbf{y}$

3. Set $\boldsymbol{\theta}^{(r)} \leftarrow \mathbf{L}^{(r)}\boldsymbol{\beta}_k$ for $r = 1, \ldots, R$ and compute

$$\mathbf{D}^{(r)} \leftarrow \text{diag}\left(g_r'\left(|\theta_1^{(r)}|\right)/\left(\varepsilon + |\theta_1^{(r)}|\right), \ldots, g_r'\left(|\theta_{N_r}^{(r)}|\right)/\left(\varepsilon + |\theta_{N_r}^{(r)}|\right)\right)$$

4. If $k = 1$, then set $M \leftarrow \max\{g_r'(0_+)\}$ and $\varepsilon \leftarrow \frac{\tau}{2RM}\min\left\{|\theta_i^{(r)}| : \theta_i^{(r)} \neq 0\right\}$

5. Set $\boldsymbol{\Omega} \leftarrow \sum \lambda_r \mathbf{L}^{(r)T}\mathbf{D}^{(r)}\mathbf{L}^{(r)}$ and compute $\boldsymbol{\delta} \leftarrow -\mathbf{X}^T\mathbf{y} + (\mathbf{X}^T\mathbf{X} + \boldsymbol{\Omega})\boldsymbol{\beta}_k$

6. If $|\delta_j| < \tau/2$ for all $j \in \{1, \ldots, p\}$ such that $|\beta_j| \geq \varepsilon$, then **goto** Step 8.

7. If $k <$ MaxIter, then **goto** Step 2.

8. Stopping criterion: if convergence is reached then the solution is $\hat{\boldsymbol{\beta}} \leftarrow \boldsymbol{\beta}_k$.

This algorithm depends on the regularization parameters $\lambda_1, \ldots, \lambda_R$, which can be chosen from a given grid of values or from an automatically determined range according to the singular values of **X**. The selection of the 'optimal values' for these parameters is a process that will not be considered here in detail. This is usually done by minimizing information criteria such as Akaike (AIC), Bayesian (BIC) or the generalized cross-validation (GCV) function. For this purpose, it is necessary to compute the degrees of freedom, which can be estimated as proposed in Hunter and Li (Hunter and Li (2005)). In order to avoid the selection of optimal parameters in an R-dimensional grid, we prefer to set $\lambda_r = \lambda \mu_r$ and set ad hoc values for the proportions $\mu_r > 0$, for $\forall r = 1, \ldots, R$, (such that $\sum \mu_r = 1$), which represent prior assumptions about relative penalty contributions and allow simplifying the process to estimating only the overall weight as a single parameter $\lambda$.

## 3. AMNR technique for MPLS methods

Although the MNR algorithm allows the implementation of a wide range of MPLS methods, it produces very small coefficients that should be estimated as zero in sparse scenarios, similarly as it happens with classical procedures (Hunter and Lange (2004)). Variable selection and active set algorithms overcome this limitation by doing feature selection and estimation simultaneously, which implicitly guarantees a higher degree of sparsity in the solution. As noticed by Mørup and others, the LARS algorithm can be stated as the iterative application of the Newton-Raphson technique over the space of selected predictors, considering a particular selection of the descent step (Mørup et al. (2008), Hastie et al. (2009)). This strategy has been used to produce the optimal estimators for LASSO and other specific models, avoiding the explicit use (and estimation) of a regularization parameter. In this section, we introduce a technique based on the use of MNR over the active set of salient features (AMNR) which generalizes this type of algorithm to deal with MPLS models including LARS as a particular case.

Let's consider the MPLS optimization problem defined in the unconstrained variant as shown in equation 1.1, where $\Psi(\boldsymbol{\beta})$ is a sum of convex functions, which guarantees the convexity of the objective or cost function $f(\boldsymbol{\beta})$. Let's now assume we are at step $k$ of the minimization of $f(\boldsymbol{\beta})$ with coefficients vector $\boldsymbol{\beta}_k$ that will be updated as $\boldsymbol{\beta}_{k+1} = \boldsymbol{\beta}_k + \mathbf{b}_k$. If we take into account that the effect of the previous steps can be absorbed by the residuals: $\mathbf{r}_k = \mathbf{y} - \mathbf{X}\boldsymbol{\beta}_k$, the cost function at step $k+1$ can be expressed as a function of the vector update $\mathbf{b}_k$:

$$f_{k+1}(\mathbf{b}_k) = \|\mathbf{r}_k - \mathbf{X}\mathbf{b}_k\|_2^2 + \Psi(\boldsymbol{\beta}_k + \mathbf{b}_k)$$

The change in the cost function from the previous iteration $\left(f_k = f(\boldsymbol{\beta}_k) = f_{k+1}(\mathbf{b}_k)|_{\mathbf{b}_k=0}\right)$ to the next iteration $(f_{k+1}(\mathbf{b}_k) = f(\boldsymbol{\beta}_k + \mathbf{b}_k))$, can be found from the second term in the first-order Taylor approximation of $f_{k+1}(\mathbf{b}_k)$ around $\mathbf{b}_k = 0$:

$$f_{k+1}(\mathbf{b}_k) \approx f_k - \left(2\,\mathbf{x}^T \mathbf{r}_k - \nabla_{\mathrm{b}}\Psi(\boldsymbol{\beta}_k + \mathbf{b}_k)|_{\mathbf{b}_k=0}\right)^T \mathbf{b}_k \qquad (3.1)$$

Interestingly, the change in the cost function, for small update vectors, is proportional to the product between the residuals and the predictors (columns of $\mathbf{X}$) minus the gradient of the penalization term (with respect to the update vector $\mathbf{b}$) evaluated at $\mathbf{b}_k = 0$, and convergence will be achieved when these terms become equal. From this equation, we can then follow a procedure similar to that used to derive the LARS algorithm (Efron (2004)), which consists in finding the update (size and direction) $\mathbf{b}_k$ that ensures the minimization of the cost function ($f_{k+1} < f_k$) by involving only one nonzero coefficient in each iteration. The set of nonzero coefficients obtained in this way is known as the active set and represented as $\mathcal{A}$, while the columns of the corresponding predictors form the matrix $\mathbf{X}_{\mathcal{A}}$.

In the more general AMNR formulation, we establish the following LARS-type constraint to be fulfilled in every iteration $k$:

$$\left|2\mathbf{X}_{\mathcal{A}}^{\mathrm{T}} \mathbf{r}_k - \nabla_{\mathrm{b}}\Psi(\mathbf{b}_k)\right| = C_{max}\mathbf{1}_{\mathcal{A}} \text{ with } C_{max} > 0, \qquad (3.2)$$

where $\mathbf{1}_{\mathcal{A}}$ represents a vector of ones with length equal to the cardinality of the active set. The common use of the absolute value in this condition in LARS aims at controlling the sign of the updated coefficients (avoiding the sign flip of the coefficients included in the active set) and ensuring that the last term in equation (3.1) is positive (Efron (2004)). We will keep this approach for the AMNR implemented here, although future versions of the algorithm might be able to avoid such trick and allow flipping signs in a natural way. Note that equation (3.2) reduces to the equivalent LARS-LASSO constraint for $\Psi$ being the L1 norm of $\boldsymbol{\beta}$, where the gradient becomes a constant and then predictors are included in the active set in such a way that they have the same correlation with the residuals (equi-angular condition).

In the general case of the AMNR, the condition is more complicated and depends on the difference between the correlation with the residuals and the gradient of the penalty function. The update $\mathbf{b}_k$ is taken on the Newton-Raphson direction, which corresponds to the direction of the OLS solution ($\mathbf{b}_k = \alpha_k(\mathbf{X}_{\mathcal{A}}^T\mathbf{X}_{\mathcal{A}})^{-1}\mathbf{X}_{\mathcal{A}}^T\mathbf{r}_k$) in the case of LARS, but corresponds to the penalized solution, i.e. the MNR solution ($\mathbf{b}_k = \alpha_k(\mathbf{X}_{\mathcal{A}}^T\mathbf{X}_{\mathcal{A}} + \mathbf{\Omega})^{-1}\mathbf{X}_{\mathcal{A}}^T\mathbf{r}_k$; with $\mathbf{\Omega}$ defined as in Step 5 of Algorithm 1) in the case of the AMNR. In both cases, the algorithm is then reduced to find only the step size of the update ($\alpha_k$) in each iteration, such that the variable selected to be included in the active set will ensure that equation 3.2 holds for every iteration. Obviously, $C_{max}$ in equation 3.2 is different for each iteration and will be ideally zero when convergence is reached. As detailed in Section 2 of the Supplementary Material (online), the step sizes for including positive ($\alpha^+$) or negative ($\alpha^-$) coefficients in the active set, are different. This allows us to impose nonnegativity (nonpositivity) constraints in a natural way by just fitting positive (negative) coefficients and leaving the rest out of the active set. Moreover, in every iteration, the minimum step size needed for a sign flip in any of the active coefficients ($\alpha^0$) is also computed, and the final step size is taken as the minimum of the three, for avoiding the flip in signs of the estimated coefficients.

Although this formulation leads us to a very flexible algorithm that can deal with general MPLS models, in this work we implement the simplest versions of the algorithm which correspond to the LASSO and Adaptive LASSO models. Similar to LARS, the AMNR algorithm can also be conveniently established on the constrained equivalent formulation for Adaptive LASSO:

$$minimize\|\mathbf{y} - \mathbf{X}\boldsymbol{\beta}\|_2^2 \text{ subject to } \sum_{j=1}^{p}\gamma_j|\beta_j| \leq \tau,$$

where $\gamma_j$ are positive weights, thus reducing this model to LASSO when they are all set to 1. This formulation avoids the explicit use of regularization parameters by replacing them by a thresholding parameter ($\tau$). However, future developments in the algebra of the AMNR approach could potentially lead to find an explicit relationship between the path of solutions of the algorithm and the regularization parameters, which might provide other conditions for estimating their optimal values. This formulation of the Adaptive LASSO model is especially relevant as many convex MPLS models are combinations of penalty functions based on L1 and L2 norms. Hence, using the trick of data augmentation (Hebiri and van de Geer (2011)), we can include the L2-based penalty terms into the data fitting term of the objective function, which implies using a value for the explicit regularization parameter defined ad hoc or computed by minimizing some information criteria for the corresponding MNR model before applying the AMNR algorithm. In this sense, many combined models reduce to a LASSO-type model, as is the case of Elastic Net (Zou and Hastie (2005)) and Smooth LASSO (Hebiri and van de Geer (2011)). Additionally, these simple models allow us to use the AMNR for handling novel models, such as the extension of the Garrote to the $p \ll n$ case and of other models to sign constrained versions, as we will show in the next sections. In Section 2 of the Supplementary Material (online), we give the AMNR derivation for the Adaptive LASSO model and the pseudo algorithm implemented. In Section 3 of the Supplementary Material (online), we demonstrate that the AMNR fulfills the optimality conditions.

## 3.1 Addressing new models with AMNR

The capability for establishing nonnegative and non-positive constraints with the AMNR algorithm can be exploited to implement the nonnegative Garrote (NNG) method (Breiman (1995)), which is stated as:

$$\widehat{\mathbf{W}} = \underset{\mathbf{w}}{\operatorname{argmin}} \left\{ (1/2) \left\| \mathbf{y} - \sum_{j=1}^{p} \mathbf{X}_j \beta_j^{ols} w_j \right\|_2^2 + \lambda \sum_{j=1}^{p} w_j \right\}; \text{with } w_j = \left( \beta_j / \beta_j^{ols} \right) \geq 0$$

where $\beta_j^{ols}$ is the j-th component of the Ordinary Least Squares (OLS) estimator. This is analogous to the following formulation:

$$\widehat{\boldsymbol{\beta}} = \underset{\boldsymbol{\beta}}{\operatorname{argmin}} \left\{ (1/2) \| \mathbf{y} - \mathbf{X}\boldsymbol{\beta} \|_2^2 + \lambda \sum_{j=1}^{p} (1/|\beta_j^{ols}|) |\beta_j| \right\}$$

This can be seen as a version of the Adaptive LASSO (ALASSO) model (see Table S1 of the Supplementary Material (online)) with sign constraints over $\boldsymbol{\beta}$ and with weights defined as $\gamma_j = 1/|\beta_j^{ols}|$. As originally conceived, the NNG is limited to $p < n$ situations because it depends heavily on the OLS estimator. However, it can be extended to the $p \gg n$ scenario if we consider making it dependent on other estimators and stating a general approach. In other words, we can use any reference estimator (for example, the LASSO, Fusion LASSO, ENET or SLASSO) and denote it as $\boldsymbol{\beta}^{ref}$. The NNG extension to the $p \gg n$ scenario is then a nonnegative version of an ALASSO model where the weights are defined from other reference solutions previously known (or computed). In this sense, it is clear that an AMNR algorithm can be designed to implement the NNG method for general $p \gg n$ conditions. It is evident that using a sparse $\boldsymbol{\beta}^{ref}$ would be helpful to promote sparsity since those high $\beta_i^{ref}$ will imply smaller penalization and those close to zero will push the corresponding variables of the NNG solution to zero. The AMNR algorithms for the ALASSO is shown in Section 2 of the Supplementary Material (online), which can be used for computing the NNG by just using the weights as defined by the $\boldsymbol{\beta}^{ref}$ and imposing the desired sign constraint when selecting the step size in the algorithm.

Another extension that we consider here is the inclusion of other penalty terms in the NNG model (i.e., extend it to MPLS models). The simplest option is just to add a quadratic term for imposing some degree of smoothness:

$$\widehat{\boldsymbol{\beta}} = \underset{\boldsymbol{\beta}}{\mathrm{argmin}}\{(1/2)\|\mathbf{y} - \mathbf{X}\boldsymbol{\beta}\|_2^2 + (\lambda_{sm}/2)\|\mathbf{L}\boldsymbol{\beta}\|_2^2 + \lambda_{sp}\sum_{j=1}^{p}(1/|\beta_j^{ref}|)|\beta_j|\}$$

which is equivalent to:

$$\widehat{\mathbf{W}} = \underset{\mathbf{w}}{\mathrm{argmin}}\left\{(1/2)\left\|\mathbf{y} - \sum_{j=1}^{p}\mathbf{X}_j\beta_j^{ref}w_j\right\|_2^2 + (\lambda_{sm}/2)\|\tilde{\mathbf{L}}\mathbf{w}\|_2^2 + \lambda_{sp}\sum_{j=1}^{p}w_j\right\}; \text{s.t. } w_j \geq 0 \quad (3.3)$$

where $\tilde{\mathbf{L}} = \mathbf{L}\mathbf{D}$, $\mathbf{D} = \mathrm{diag}(\beta_1^{ref}, \beta_2^{ref}, \ldots, \beta_p^{ref})$ and $\mathbf{L}$ is a structure matrix that can be set to the identity matrix (i.e. implying independence of $w_j$) or to any other matrix (e.g. first or second difference operators). This extension of the NNG (equation 3.3) with a smoothness term will be called the Smooth Nonnegative Garrote (SNNG). In practice, the solution is found by $\hat{\beta}_j = \widehat{w}_j\beta_j^{ref}$, avoiding the division by $|\beta_j^{ref}|$ (in any step of the algorithm) which is very important when $\boldsymbol{\beta}^{ref}$ is a sparse solution. In that case, the condition $w_j \geq 0$ implies that elements that are zero in the $\boldsymbol{\beta}^{ref}$ will also be zero in $\boldsymbol{\beta}$, making the problem smaller and the estimation faster as we only need to re-estimate coefficients that are nonzero in $\boldsymbol{\beta}^{ref}$ and also making the SNNG solution sparser than the one used as reference.

On the other hand, equation (3.3) can also be seen as a Smooth LASSO model (Hebiri and van de Geer (2011)) with sign constraints if we set $\beta_j^{ref} = 1$, $\forall j$, and take $\mathbf{L}$ as the first differences matrix. We will call this extension of Smooth LASSO with nonnegative constraints, as NN-SLASSO. To our knowledge, the SLASSO model has not been treated with nonnegative restrictions, we will then explore its performance to solve ill-posed problems in this study. Other extensions that will not be explored here can be easily derived from the more general model, such as nonnegative versions of ENET and Ridge L (i.e., using $\lambda_{sp} = 0$ ).

## 4. Performance of MNR and AMNR: simulation study

The goal of this simulation study is to investigate the performances of MNR and AMNR algorithms for different penalized models. To this end, we generate 100 independent samples of the simulation design for three different $n/p$ relations ($n/p = \{0.05, 0.25, 0.5\}$), corresponding to 200 predictors ($p = 200$) and n observations ($n = \{10, 50, 100\}$), respectively. The simulation design consists in a solution with three active regions (nonzero components) that will be called the 'bell', 'square' and 'point' sources, and the use of the linear model $\mathbf{y} = \mathbf{X}\boldsymbol{\beta} + \boldsymbol{\varepsilon}$ where $\beta_j = 0$ except for:

$$\beta_j = \begin{cases} e^{-0.015(j-50)^2}, & \text{for } 30 < j < 70 \text{ (\textbf{bell})} \\ 1, & \begin{cases} \text{for } 95 < j < 105 \text{ (\textbf{square})} \\ \text{and } j = 150 \text{ (\textbf{point})} \end{cases} \end{cases}$$

The components of $\mathbf{X}_j$ and $\boldsymbol{\varepsilon}$ are standard normal, which leads to a theoretical SNR of about 13 db. For this data, we estimate $\boldsymbol{\beta}$ with different penalty methods and algorithms. This simulated solution is difficult to recover with any of the methods, as it is a piece-wise combination of smooth, constant and isolated coefficients. It is therefore appropriate to evaluate the flexibility of the complex models based on combinations of different penalty functions. To evaluate the performances of the different algorithms, we used three of the quality measures that have been used and described in the literature (Obuchowski et al. (2018)). Area Under the ROC Curve (AUC), Relative Error ($RE = \sum_{i=1}^{n}(\beta_i - \hat{\beta}_i)^2 / \sum_{i=1}^{n}\beta_i^2$), and the computational time (for one solution). In the analysis of simulated data, we will report boxplots of these measures computed using 100 repetitions corresponding to different noise instances.

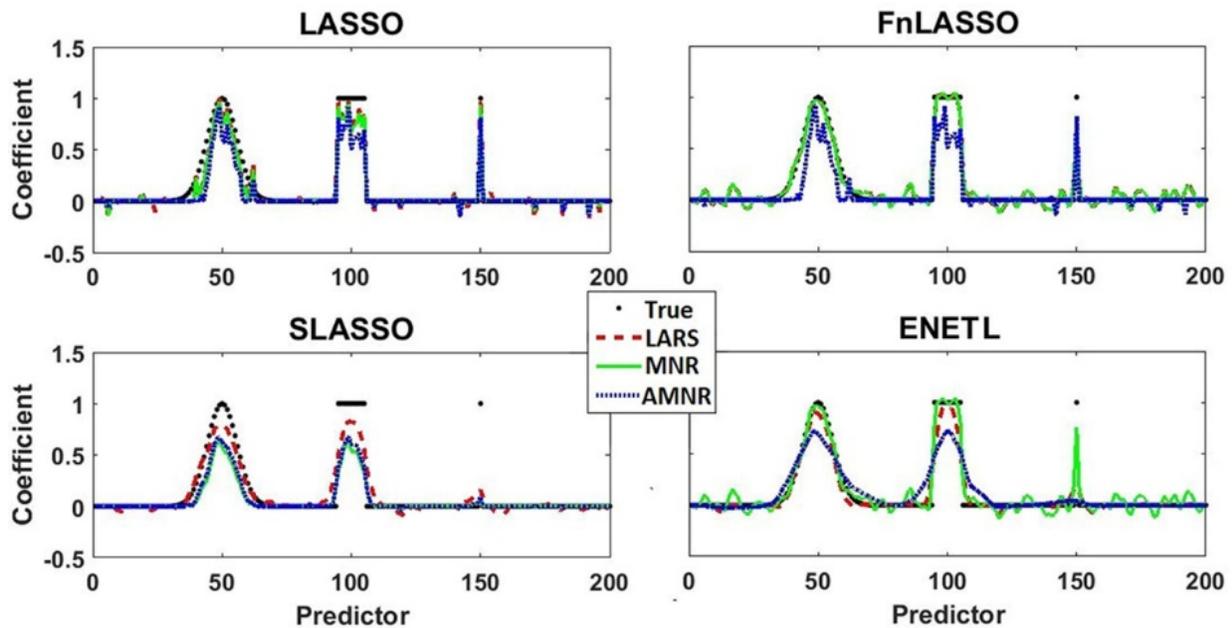

**Figure 1:** Simulated example with $n = 100$ samples and $p = 200$ predictors for testing the LASSO, Fusion LASSO (FnLASSO), Smooth LASSO (SLASSO), and Elastic Net (ENET L) models. Dots represent the true simulated coefficients, the dashed lines correspond to the solutions estimated by LARS algorithm, the solid lines represent those estimated by MNR and the dotted lines to those estimated by AMNR.

Figure 1 and Figure 2 illustrate an application of the two proposed algorithms for different models. Both figures show a simulation with $n = 100$ samples and $p = 200$ predictors ($n/p = 0.5$), and the solutions corresponding to those with the median AUC (i.e., the repetition whose AUC is the closest to the median of all AUC). In Figure 1, we compare the LASSO, FnLASSO and SLASSO solutions obtained by using the well-known LARS algorithm as well as with the MNR and AMNR algorithms presented in this article. Generally, the MNR and the AMNR offered solutions with similar behavior as those offered by LARS. In the ENET L and FnLASSO, MNR provides the least sparse solutions but reconstructing better all sources (bell, square, point). AMNR seems to estimate over-sparse solutions, missing the point source in the case of SLASSO and ENET L models.

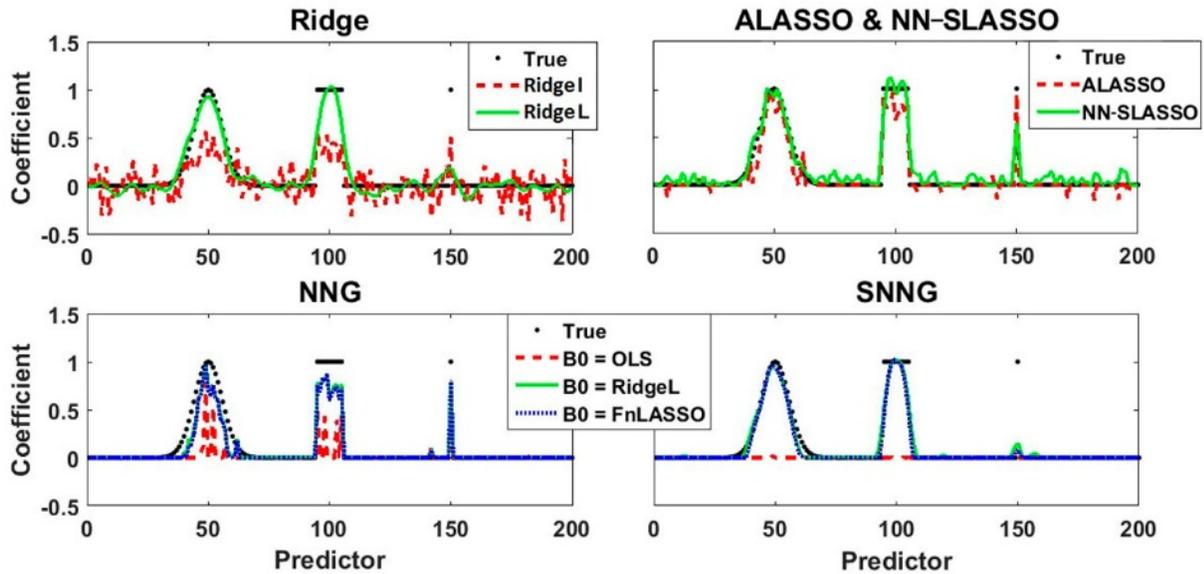

**Figure 2:** Simulated example with $n = 100$ samples and $p = 200$ predictors for testing the performance of the AMNR algorithm for different models: Ridge, Adaptive LASSO (ALASSO), NonNegative Smooth LASSO (NN-SLASSO), NonNegative Garrote (NNG) and Smooth NonNegative Garrote (SNNG). Dots represent the true simulated coefficients in all panels. For the Ridge model, a dashed line represents the classical solution, and the solid line represents the solution using the matrix of second differences (**L**). For NNG and SNNG models, we used different reference solutions (B0): dashed, solid and dotted lines correspond to using the ordinary least squares (OLS), Ridge with Laplacian operator (Ridge L) and Fusion LASSO (FnLASSO), respectively.

Figure 2 shows the estimators for Ridge (with and without using the Laplacian operator), ALASSO, NN-SLASSO, NNG and SNNG penalized models, obtained by using the AMNR algorithm. In general, the use of nonnegativity constraints in the NN-SLASSO, or the use of a reference estimator in NNG and SNNG, allows obtaining sparser estimators, more similar to the true simulation. In all cases, the estimator corresponding to the point source suffered from the insufficient data problem and was over shrunk when using smoothness constraint. In particular, the SNNG estimator behaves as a smooth though sparser version of the reference estimator: the most salient features are enhanced while the smaller are discarded. On the contrary, the NNG solution enhances isolated sources at the cost of degrading the reconstruction of the smooth patches. ALASSO and NN-SLASSO offered good reconstruction of the three regions but with many (small) spurious nonzero coefficients.

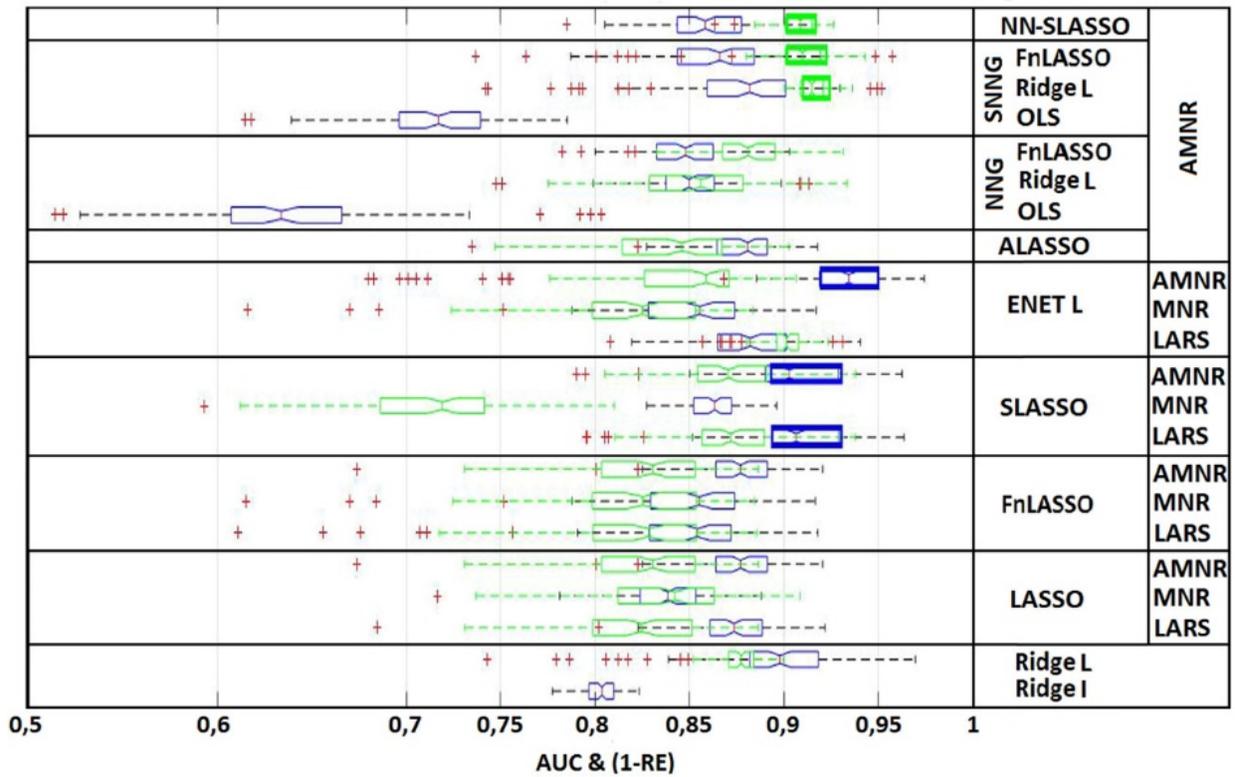

**Figure 3:** Boxplots of the Area Under Curve (AUC, in dark gray) and one minus the Relative Error (1-RE, in light grey) for the 100 solutions estimated with each method (combining model and algorithm), from the simulations using 200 predictors ($p = 200$) and 100 observations ($n = 100$). The Ridge solutions were computed with the classical regularized Tikhonov inverse solution (Tikhonov et al. 1995).

For each of the 100 simulations appearing in Figure 1 and Figure 2, for the nine different models and three algorithms, we computed the three measures for a quantitative evaluation of the quality of the reconstruction. Figure 3 presents a boxplot of the AUC and (1-RE) together, for an easier interpretation of results. In both cases we only show the results that are above 0.5. We can see that most of the methods provided medians of AUC and 1-RE above 0.8, but none of them had both measures over 0.9. Only SLASSO (with LARS and AMNR) and ENET L (with AMNR) had median AUC over 0.9 while the lowest RE (<0.1) was given by NN-SLASSO and SNNG (with AMNR) using Ridge L and FnLASSO as reference solutions. Note that Ridge L also offers good performance in terms of AUC and RE in this case with high $n/p$ ratio, but this does not hold for other lower ratios as will be shown next. Figure S1 of the Supplementary Material presents a boxplot of the computational time (in seconds), showing that -besides the non-iterative Ridge solutions- the faster models are LASSO using LARS algorithm and NN-SLASSO and NNG using

AMNR for whatever reference estimator. As expected, the same models using AMNR were generally faster than when computed with the MNR algorithm.

| No | Methods (B0) | Algorithm | Mean for three n/p ratios | | |
|---|---|---|---|---|---|
| | | | RE | AUC | TIME |
| 1 | Ridge I | Tikhonov | 0.7547 | 0.6982 | **0.0103** |
| 2 | Ridge L | | 0.6341 | 0.7996 | **0.0110** |
| 3 | LASSO | LARS | 0.7953 | 0.7984 | 0.4053 |
| 4 | FnLASSO | | 0.6892 | 0.7718 | 1.0905 |
| 5 | SLASSO ▲ | | **0.3416** | **0.8581** | 1.3002 |
| 6 | ENET L ▲ | | **0.3587** | 0.7861 | 1.2328 |
| 7 | LASSO | MNR | 0.7568 | 0.7074 | 5.5728 |
| 8 | FnLASSO | | 0.4861 | 0.7760 | 3.4986 |
| 9 | SLASSO | | 0.5103 | 0.8104 | 1.4545 |
| 10 | ENET L | | 0.4769 | 0.7759 | **0.1001** |
| 11 | LASSO | AMNR | 0.7706 | 0.8038 | 0.6737 |
| 12 | FnLASSO | | 0.7706 | 0.8038 | 0.8614 |
| 13 | SLASSO ▲ | | **0.3531** | **0.8522** | 1.8470 |
| 14 | ENET L ▲ | | **0.3295** | **0.8658** | 1.3819 |
| 15 | NNG (B0 = OLS) | | 1.4181 | 0.6839 | 0.3558 |
| 16 | NNG (B0 = Ridge L) | | 0.7110 | 0.7895 | 0.3769 |
| 17 | NNG (B0 = FnLASSO) | | 0.7039 | 0.7872 | **0.3226** |
| 18 | SNNG (B0 = OLS) | | 0.9749 | 0.6521 | 0.9339 |
| 19 | SNNG (B0 = Ridge L) | | 0.5996 | 0.7144 | 2.1342 |
| 20 | FnLASSO ▲ | | **0.3911** | **0.8366** | 2.1193 |
| 21 | ALASSO | | 0.7766 | 0.8003 | 0.5800 |
| 22 | NN-SLASSO ▲ | | **0.3378** | **0.8179** | **0.2127** |

**Table 1:** Mean of quantitative quality measures (relative error, area under the curve and computation time) across all n/p ratios for all combinations of models and algorithms. In the case of the NNG and SNNG, **B0** represents the reference solution. The best 5 methods in each column were highlighted and the overall best 5 methods were marked with a black triangle.

Figure S2 of the Supplementary Material (online) shows the median of AUC, median Relative Error and median of the computation time across the 100 estimated solutions in the three cases of $n/p$ ratios {0.05, 0.25, 0.5}. These results are summarized in Table 1 showing the mean across the three $n/p$ ratios for all estimated models and algorithms. The methods with better performance for any $n/p$ ratio, were SLASSO (with LARS and AMNR algorithms), ENET L (with LARS and AMNR algorithms), SNNG (with FnLASSO as reference estimator) and the nonnegative version of SLASSO (NN-SLASSO). These results are consistent with the qualitative pictures given in Figure 1 and Figure 2, where SLASSO, ENET L, SNNG and NN-SLASSO better reconstructed the simulated bell and square regions. It is clear that the general difficulties in reconstructing the point source will not be largely reflected in the quantitative measures, as it is just one out of 200 estimated points. The analysis of the time necessary for computing one solution showed that the fastest methods are Ridge I and Ridge L, NNG (with the three reference solutions) and NN-SLASSO.

## 5.   MNR and AMNR algorithms for solving the EEG inverse problem

Since the last decade of the past century, much effort has been devoted to the development of methods for EEG/MEG source imaging, i.e., for identifying the generators of the EEG/MEG, which is also known as the EEG/MEG Inverse Problem (EEG IP). Mathematically, this is an ill-posed problem and finding a solution requires the use of additional or prior information about the properties of the sources. Therefore, the EEG IP is usually established as a penalized regression model (Pascual-Marqui et al. (1994); Dale et al. (2000)). However, there is currently no ground truth available about which electrophysiological sources are active in real EEG/MEG experiments. Therefore, the problem of finding the best inverse solution from the many methods proposed (Grech et al. (2008)) is not straightforward. In this context, we have followed the strategy to propose very flexible models that can adjust solutions to the data at hand (Valdés-Sosa et al. (2006)). We have indeed proposed to formalize this problem as a more general MPLS model and have previously studied the performance of inverse solutions obtained from models such as LASSO, FnLASSO and ENET, using LQA and MM in simulated and real EEG data (Vega-Hernández et al. (2008)).

In this section, we explore the use of the AMNR algorithm for solving the EEG inverse problem with multiple penalties. This will allow us to compare the behavior of the recently proposed algorithm with previous MNR techniques in such a difficult problem. In addition, the AMNR will allow us to try models with combination of smoothness/sparsity and sign constraints that have never been applied to the EEG IP before. Simulated and real data were used for a preliminary comparison with the equivalent solutions using the MNR algorithm in terms of quality measures (localization error and blurring) (Vega-Hernández et al. (2008)).

## 5.1 Simulated data

The synthetic data consisted in four different sets of simulated primary current density (PCD) distributions, all of them simulated as a three-dimensional Gaussian source with amplitude of 10 nA/mm$^2$ and width of 10 mm (spherical). Each set contain seven PCDs: a 'centroid' PCD with maximum located in a particular anatomical structure of a brain space of 3862 generators, and 6 others derived from this one by locating the maxima in each of the 6 closest neighbor generators. Maximum values of the simulated PCDs were in 1) the cingulate region left (Cingulate), 2) occipital pole left (Occipital), 3) postcentral gyrus (Postcentral), and temporal gyrus right (Temporal) as shown in the first row of Figure S3 of the Supplementary Material (online). Talairach Coordinates (Talairach and P. Tournoux, (1988)) of the maximum value of each simulated PCD appear in Table S2 of the Supplementary Material (online).

The design matrix (known as the Electric Lead Field) for this brain was computed for an array of 19 electrodes from the 10/20 system using a three-spheres pricewise homogenous and isotropic head model (Riera, (1999)). The simulated voltages were obtained through the equation:

$$\mathbf{V}_{(Ne \times 1)} = \mathbf{K}_{(Ne \times 3 \cdot Ng)} \, \mathbf{j}_{(3 \cdot Ng \times 1)} + \boldsymbol{\varepsilon}_{(Ne \times 1)},$$

where $\mathbf{j}$ is the simulated PCDs, $\mathbf{K}$ the lead field and $\mathbf{V}$ is the vector of electric potentials (i.e., simulated EEG as if it was measured on an array of electrodes distributed on the scalp surface). Additive white noise was set up in order to have a signal-to-noise ratio (SNR) of 14.8 db. $Ne$ represents the number of electrodes (19) and $Ng$ the number of sources or generators (3862), i.e., the number of grid points obtained from the discretization of the source space inside the brain. Note that as the PCD in each source is a vector magnitude, the solution **j** has $3Ng$ elements corresponding to coordinates x, y, z of the PCD in each source, effectively estimating not only the amplitude but the orientation of the PCD in each voxel.

| No | Methods (B0) | Algorithm | Accuracy | Normalized Localization Error | Normalized Blurring |
|---|---|---|---|---|---|
| 1 | **Ridge I** | Tikhonov | 0.439 ± 0.075 | 0.656 ± 0.215 | 0.022 ±0.049 |
| 2 | **Ridge L** | | 0.442 ± 0.069 | 0.654 ± 0.164 | 0.009 ±0.009 |
| 3 | **LASSO** | LARS | 0.951 ± 0.106 | 0.629 ± 0.214 | 0.798 ±0.198 |
| 4 | **FnLASSO** | | 0.464 ± 0.075 | 0.588 ± 0.242 | 0.802 ±0.124 |
| 5 | **SLASSO** | | **0.968 ± 0.006** | 0.669 ± 0.167 | 0.486 ±0.289 |
| 6 | **ENET L** | | **0.968 ± 0.006** | 0.697 ± 0.159 | 0.454 ±0.290 |
| 7 | **LASSO** | MNR | 0.458 ± 0.100 | 0.635 ± 0.179 | 0.842 ±0.113 |
| 8 | **FnLASSO** | | 0.444 ± 0.065 | 0.520 ± 0.239 | 0.717 ±0.134 |
| 9 | **SLASSO** | | 0.828 ± 0.107 | 0.653 ± 0.209 | 0.035 ±0.164 |
| 10 | **ENET L** | | 0.862 ± 0.116 | 0.522 ± 0.239 | 0.698 ±0.185 |
| 11 | **LASSO** | AMNR | **0.952 ± 0.101** | 0.652 ± 0.186 | 0.821 ±0.195 |
| 12 | **FnLASSO** | | **0.952 ± 0.101** | 0.572 ± 0.206 | 0.696 ±0.378 |
| 13 | **SLASSO** | | 0.949 ± 0.100 | 0.666 ± 0.164 | 0.499 ±0.290 |
| 14 | **ENET L** ▲ | | **0.952 ± 0.196** | **0.776 ± 0.181** | 0.404 ±0.349 |
| 15 | **NNG (Ridge L)** | | 0.880 ± 0.131 | 0.711 ± 0.190 | **0.869 ±0.114** |
| 16 | **NNG (FnLASSO)** | | 0.893 ± 0.128 | 0.618 ± 0.293 | **0.903 ±0.087** |
| 17 | **SNNG (Ridge L)** ▲ | | **0.952 ± 0.101** | **0.796 ± 0.153** | 0.592 ±0.437 |
| 18 | **SNNG (FnLASSO)** ▲ | | **0.953 ± 0.101** | **0.776 ± 0.180** | 0.628 ±0.423 |
| 19 | **ALASSO** | | **0.952 ± 0.101** | 0.643 ± 0.208 | 0.856 ±0.125 |
| 20 | **NN-SLASSO** ▲ | | **0.952 ± 0.101** | 0.792 ± 0.193 | **0.914 ±0.124** |

**Table 2:** Mean ± standard deviation of the accuracy, normalized Localization Error and normalized Blurring of the 28 inverse solutions for each simulated data. The three best numbers in each column are highlighted.

Using the whole set of simulations, we compared the performance of inverse solutions obtained for all models and algorithms, in terms of the accuracy of the reconstruction (Fawcett (2006)) and of normalized versions of the 'localization error' and 'blurring', as defined in (Vega-Hernández et al. (2008)). Therefore, all these three quality measures will give values close to 1 for perfect reconstructions and close to 0 for bad reconstructions. Table 2 shows the mean and standard deviation of these normalized quality measures across the 28 estimated inverse solutions.

Results showed that ENET L, SNNG (with Ridge L and FnLASSO as reference solutions) and NN-SLASSO, offered the best overall performance in reconstructing the simulated sources, all of them using the AMNR algorithm (marked with a black triangle in Table 2). In general, models computed with the AMNR offered better accuracy in the reconstruction and better localization of the maximum activation than the same models computed using MNR (LASSO, FnLASSO, SLASSO, ENET L). The ENET L, SNNG and NN-SLASSO computed by AMNR showed the best localization ability, but -among them- only the NN-SLASSO presented sources with blurring similar to that of the true simulation. Typically, sparse methods showed better estimation of the blurring, as is the case of NN-SLASSO, NNG, ALASSO and LASSO. Interestingly, both sparse and smooth methods led to solutions with high accuracy when using AMNR and LARS but not with MNR or direct computation (Ridge).

Maximum intensity projection (glass-brain visualization) of the estimated sources by the best methods according to Table 2 are shown in Figure S3 of the Supplementary Material (online), corresponding to the simulated 'centroid' PCDs in each region. We also added the Ridge L solution, which is mathematically equivalent to a classical solution known as LORETA in the field of EEG source localization (Pascual-Marqui et al (1994)). As expected, the Ridge L solutions are very smooth, while ENET L and SNNG methods (computed with AMNR) offered solutions that fluctuate between different degrees of sparsity/smoothness. Also, the use of sign constraints (allowed by AMNR) in the new inverse solutions SNNG and NN-SLASSO, led to sparser solutions than the unconstrained counterparts. SNNG solutions seem to be sparser versions of the reference solutions but without removing all ghost sources. The NN-SLASSO solutions are over-sparse but showing much less ghost sources as a convenient side effect. This solution also improves the localization of the main source with respect to ENET L, offering a very good localization even for the deepest simulated PCD (Temporal).

## 5.2 Visual event-related EEG

The real data belongs to a visual event-related experiment, explained in detail in Rodríguez (2012) (Rodríguez et al. (2012)). Briefly, the experiment consisted in presenting to the subject many trials of a sequence of visual stimuli. Each trial started with the presentation of a fixation cross for 200 ms, which was followed by a face or an image (scrambled face) for 83 ms and then immediately masked with a different scrambled image. The total combined duration of the stimulus and mask was fixed at 200 ms. Then a blank screen was presented, and participants had up to 1770 ms to make their response by pressing different keys in the keyboard. Namely, participants were instructed to rate their perception using a 4-point scale: *sure* (a face was presented), *fairly sure* (a face was presented), *possibly* (saw a face), and *no impression* (of a face). Correct identification of the presentation of a face was assumed in those trials where a face was presented, and the response was *sure* or *fairly sure*. Incorrect identification of a face was assumed in the same trials when the response was *possibly* or *no impression*.

For trials where a face is presented, the brain produces a voltage transient response (known as visual event-related potential, ERP) that can be extracted from noisy EEG recordings by averaging all the stimuli locked data (generally, 1 s-long trials or epochs extracted with respect to the stimuli onset). This ERP typically shows a negative peak around 170 ms (known as N170 component), after presentation of the stimulus. The amplitude of this peak is different for the cases whether the subject correctly recognizes a face or not, where the ERP analysis is conducted by separately averaging only the trials corresponding to each case. For each subject, the N170 amplitudes were measured as the mean voltage within a 30-ms time window centered at the peak of the component, for each condition separately. These amplitudes for all electrodes formed the topographies (maps over the scalp) that were used for source localization (i.e., they were our observed data for solving the EEG inverse problem). The sources of the N170 peak were estimated separately for the topographies corresponding to correct and incorrect responses with the use of Ridge L, the ENET L and NN-SLASSO methods computed by the AMNR algorithm.

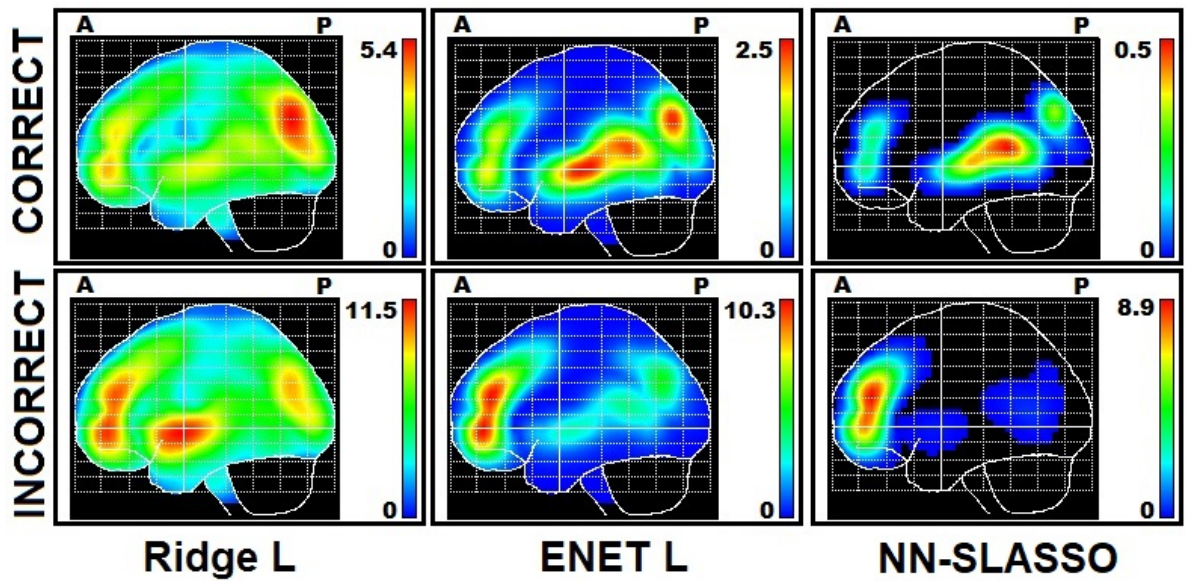

**Figure 4:** Sources of the evoked N170 peak corresponding to topographic maps of correct face detection (upper row) and incorrect detection (bottom row). Maximum intensity projection in the sagittal plane is shown and the amplitudes of the sources are coded in the corresponding color bar to the right of each panel. A and P stand for Anterior and Posterior parts of the brain.

Figure 4 shows the maximum intensity projection in the sagittal plane of the estimated sources of the N170 peak, for each condition: correct responses (top row) and incorrect responses (bottom row). Sources of the N170 for correct responses were found in the superior temporal gyrus (left and right), in the middle frontal gyrus (left and right) and in the right middle occipital gyrus. Ridge L and ENET L showed less activation on the occipital region for incorrect detection than for the correct detection, while NN-SLASSO offered a sparser solution with no occipital sources in the incorrect condition. On the contrary, the three methods showed stronger activations in the frontal areas in the case of incorrect detection as compared with sources for correct detection. In general, NN-SLASSO showed a cleaner picture than ENET L, while Ridge L gave an over smoothed solution with other confusing activations.

## 5.3 Resting-state EEG study in active older adults.

Walking speed (or gait speed (GS)) is used in the clinical practice as the main predictor of adverse outcomes and is considered a sign of elderly health. It has been linked to disability, hospitalization and death. Walking speed is primarily applied to monitor the functional capacity of older adults and forecast their rate of age-related decline. Recently, there has been an increase in the amount of studies of gait speed as a predictor of a decline in the brain function in older adults. (Rosano et al, (2012); Varma et al, (2016); Pinter et al. (2017)). Main findings suggest that there is an increased brain activation in the prefrontal cortex in response to cognitive tasks during the gait. Because of the complex cognitive processes involved in the speed of the gait, a hypothesis has emerged which states that the slowing down of the motor functions could be an early and sensitive indicator of cognitive sub-clinical deficits in cognitively normal individuals, whereas mobility decline and slow gait predict cognitive deterioration and progression to dementia (Mielke et al. (2012)).

In this section we explore the sources of the resting-state EEG measured on a cohort of elders in two different times, to evaluate the correlation with their cognitive decline. This real data is a subset of a prospective study ran in 2010 and 2016, explained in detail in García-Agustín et al (2020). The original study involved 90 community-dwelling participants over 60 years old that regularly practiced mild-to-moderate exercise in their communities. For this subset, participants were divided into two groups according to their gait speed (GS) as measured in the 2010 and 2016 evaluations. GS was quantified from measuring the time spent to cover 4 meters at a normal pace and was expressed in meters per second (m/s). The main interest was put in two groups defined as follows: Group GG, 15 participants with preserved GS (i.e., >0.8 m/sec) in both 2010 and 2016 evaluations; Group BB, 15 participants with abnormal GS (<0.8 m/sec) in both evaluations.

Figure 5 shows the difference between the estimated sources obtained with the Ridge L and NN-SLASSO solutions from the resting state EEG measurements in 2016 and 2010, represented as cortical generators of the theta band at the 6.25 Hz frequency, for each of the two groups studied. The maximum activations in each group were found in the frontal inferior bilateral areas, inferior temporal and occipital regions. Although the two types of solutions have different scales, both show that the increase in the energy of these generators in the BB group is greater than in the GG group. In general, NN-SLASSO showed sparser solutions than those of Ridge L, which might help to better interpret all generators, or discard those that are likely to be spurious sources.

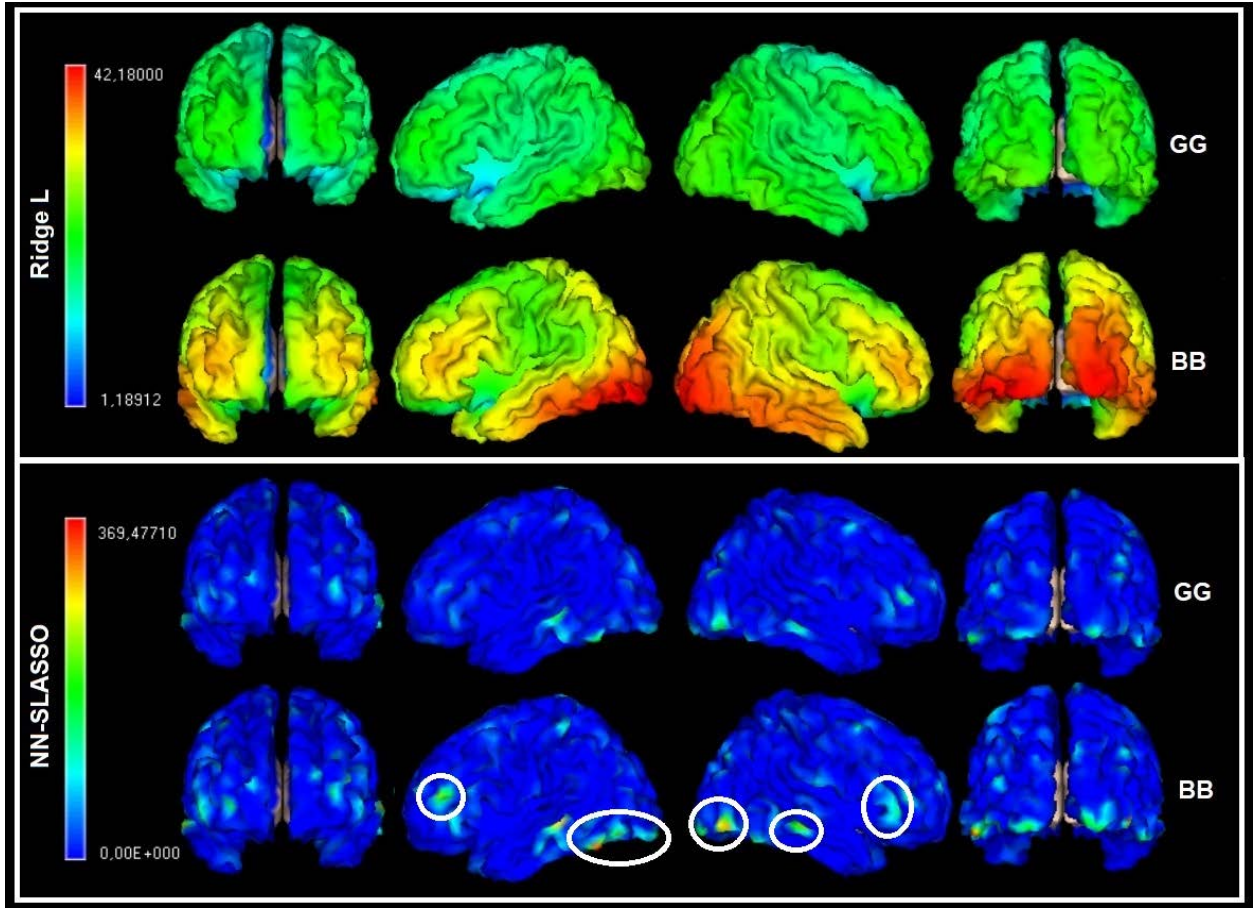

**Figure 5:** Surface 3D representation of the difference of cortical generators obtained from 2016 and 2010 resting-state EEG measurements, with the RidgeL and NN-SLASSO inverse solutions in the two groups studied (GG and BB). Sources of Theta band (at 7.03 Hz) in the frontal, left, right and posterior views (from left to right panels) of the brain. The amplitudes of the sources are coded in the corresponding color bar to the left of each panel.

## 6. Discussion

### 6.1 New algorithms for multiple penalized least squares models

In this work, we make a formal presentation of the MNR algorithm used in previous studies (Vega-Hernández et al. (2008), Sánchez-Bornot et al (2008)), and showed that the MNR could be applied for estimating general MPLS models. The main advantage is that this opens the possibility of recovering sparse and smooth estimators using combinations of L1 and L2 penalty functions. It would also provide an algorithmic framework for exploring other models, such as an extension of SCAD for estimating smooth features, which can be explored in future studies. Known techniques such as the LQA and MM algorithms can be seen as variants of our MNR technique, since these have been only adapted to implement particular models.

Although the MNR algorithm allows to implement many different penalized models, one disadvantage is that when using sparsity-promoting penalties, the estimators still give many small coefficients that should be zero, like in LQA and MM algorithms. This means that a procedure for thresholding the solutions should be included in the algorithm to recover sparse solutions, similar to the approaches implemented in MM and LQA. Another more sophisticated approach is the LQA-Fext (Sánchez-Bornot et al (2008)) which proposes to find a sub-optimal solution, but computationally feasible. To avoid regression with all variables, this procedure makes an iterative statistically selection of variables and, therefore, the final estimate is made using only a set of variables whose coefficients are nonzero. The statistical thresholding is based on the False Discovery Rate (FDR); thus, it can be said that the estimated coefficients are significantly different from zero. Unfortunately, the solutions also depend on the arbitrary value of another parameter, in this case the q-value of the FDR.

In this work we also introduced the AMNR algorithm, which is based on the application of the MNR approach restricted to a space of selected features, i.e., using the "active set" strategy. We showed that this algorithm can be applied to estimate many MPLS models and illustrated its potential application for solving them. The proposed AMNR takes advantage of using the Newton-Rapson (NR) direction in the space of active variables as the descent direction and of the possibility of reducing many MPLS models including L2-norm penalties to LASSO, or more generally, to Adaptive LASSO models. Moreover, the AMNR allows considering sign constraints in a natural way, in addition to sparsity and/or smoothness.

Although we implemented a simple version here, the proposed AMNR technique can also be regarded as a general template algorithm depending on the gradient of the penalty functions used, where only two main steps need specification: the selection of the next variable to be introduced in the active set and the calculation of the step-length $\alpha$ in the descent direction. In this general view, the LARS algorithm can be considered as a particular case of AMNR: first, the selection step includes the variable with the highest correlation with the residuals vector (in absolute value) and second, the step $\alpha$ is taken as the smallest positive value, such that some new variable joins the active set (Efron et al (2004)). Another particular case would be the forward selection method, by selecting the variable that, together with the variables in the active set, offers the lowest fitting error and then taking α = 1 for all iterations.

Using this general framework, we showed that the AMNR can also be applied to other known general nonlinear optimization problems such as Smooth LASSO or Adaptive LASSO. Specifically, the AMNR technique allowed us to propose an extension of the NonNegative Garrote (NNG) method for the $p \gg n$ scenario by using different reference solutions, which conveys the NNG advantages to this challenging scenario. Another family of new methods was also introduced by including an L2-norm penalty to the NNG model to combine sparsity, smoothness and nonnegativity constraints. These were the Smooth NNG (SNNG) and a nonnegative version of the Smooth LASSO (NN-SLASSO). Despite the flexibility of this technique, the application of AMNR to a particular model implies the derivation of a specifically tuned algorithm, which is supported by the algebraic engine that accompanies the procedure.

Similar to any penalized regression approach, another important issue for applying the MNR and AMNR algorithms is the appropriate choices of regularization parameters with respect to variable selection. The accurate estimation of these parameters and the accurate variable selection can be conflicting goals. Indeed, sometimes one gets good performances for the variable selection criteria and not so good performance for the estimation criterion (Gijbels and Vrinssen (2015)). A crucial question is also how to define an appropriate criterion for selecting the regularization parameters when both tasks, estimation and variable selection, are simultaneously addressed. This is a challenging and open research question. In this approach, we chose optimal values of the regularization parameters as those minimizing the generalized cross-validation function (GCV) (Golub, Heath and Wahba, (1979)), when needed in a step previous to the AMNR estimation. Our results suggest that the simultaneous variable selection and estimation performed in the AMNR strategy led to better reconstructions than the use of the MNR algorithm for the same models. This might be explained by the increased influence of an erroneous estimation of optimal regularization parameters by GCV in the solutions obtained with the MNR approach.

## 6.2 Validation of the AMNR algorithm

In a preliminary simulation study, we showed that the AMNR and MNR algorithms provided very similar solutions to those given by the LARS algorithm in the case of known methods like the LASSO family, but the MNR is the slowest of them. Solutions estimated from 100 independent repetitions (changing the additive noise) in three cases of $n/p$ ratio (0.5, 0.25, 0.05), showed that the sparser methods behaved better for smaller $n/p$ ratios. This can be related to the fact that when a smaller amount of data is available, stronger and more precise restrictions are needed. The SLASSO, ENET L, SNNG and NN-SLASSO offered the best reconstructions (median AUC higher than 0.9 and median relative error below 0.1, see Figure 3). The methods NN-SLASSO, ENET L and SLASSO also showed the best overall performance for any $n/p$ relation. This suggested that they are the best methods to study highly underdetermined problems such as the EEG inverse problem.

In the analysis of EEG simulated data, the solutions estimated by the AMNR algorithm showed better localization and estimation of the degree of sparsity than the solutions obtained by the MNR algorithm. Also, some of the new methods offered promising solutions to the EEG inverse problem. These models were the Smooth Nonnegative Garrote (SNNG) (using smooth solutions such as the Ridge L or FnLASSO as the reference estimator) and the Nonnegative Smooth LASSO (NN-SLASSO) method. The NN-SLASSO proved to be consistent in finding solutions with low localization error for all different groups of simulations tested, although with a tendency to provide excessively sparse distributions. In general, we found that the NNG and SNNG methods offer solutions which maintain the location of sources shown by the reference solution but with increased sparsity. This result could be exploited in cases where a rough solution with good localization is available, as it is typically the case of Ridge L, which is also fast to compute. The study of the performance of NNG and SNNG when using other more sophisticated solutions as the reference estimator should be carried out in the future.

A major point regarding the evaluation of a good estimation of EEG sources is the capability to correctly locate deep generators (that is, sources that are far from the electrodes). In most of the current methods, specifically those based on penalized regression using L2 norms, solutions are not capable to correctly locate generators in the temporal lobe or in subcortical regions such as thalamus and brainstem (Pascual-Marqui, (1999)). In our study we also observed that sources closer to the electrodes (Postcentral and Cingulate) were better located than those farther from electrodes (Temporal and Occipital) by all methods computed by the AMNR (see Figure S3 of the Supplementary Material (online)). Although the blurring of the solutions obtained varied in both cases, solutions estimated for deeper regions presented more ghost sources (estimated sources that are not present in the simulation), which usually makes harder the identification of truly activated regions. In our results, it was particularly interesting to find that the NN-SLASSO consistently showed a good location and a low number of ghost sources even for the simulated data from deeper brain regions. This suggests that more exhaustive studies should be made to validate the NN-SLASSO as a promising candidate for stable and sparse EEG source imaging.

Another interesting topic to discuss is the use of nonnegativity constraints in the context of the EEG inverse problem. In our study we found that the use of nonnegativity constraints led to sparser sources without losing real activations. The primary current density (PCD) is a vector field and therefore, an inverse method should be able to provide negative values for a proper estimation of the vector directions. In this sense, the directions obtained by the methods using nonnegativity constraints might be not reliable, and caution must be taken when interpreting results. However, there are other scenarios in which the directions of the vector field might be known or can be constrained by physiological considerations. Alternatively, we could follow a general approach in which signs of nonzero coefficients (after convergence of the sign-constrained solution) can be assigned such that they match the signs of the corresponding coefficients in a non-sign-constrained reference solution (e.g., OLS, Ridge). Particularly, this can be easily done for the NNG versions proposed here by multiplying the final sign-constrained solution by the sign of the reference solution. Future work should be devoted to a more thorough analysis of the validity and usefulness of this approach for general cases.

Finally, we performed the source localization analysis of two real experimental EEG data with some of the new methodologies in comparison with well-known methods. In a case of a visual event-related EEG from healthy young subjects, the three methods evaluated (Ridge L, ENET L and NN-SLASSO by AMNR) showed PCD distributions with main activations located in brain areas which were in accordance with previous fMRI studies showing that conscious face detection was linked to activation of fusiform and occipital face areas (see Rodríguez et al. 2012 for more details). However, both ENET L and NN-SLASSO presented sparser solutions with an easier interpretation. Although a thorough validation is needed, these results suggest that these new inverse solutions can be used for source localization analysis in other experimental data where ERPs provides relevant information on the physiological brain state. In particular, recent reviews have shown the relevance of using event-related EEG potentials for diagnosis of Alzheimer Disease (AD) (Golub et al. (1979), Cassani et al (2018) and Hedges et al. (2016)). Therefore, it will be very important to perform future studies on the ability of these methods to find differences between healthy people and AD patients in terms of the electrophysiological sources estimated when performing a cognitive task.

In a resting-state EEG study associated to walking speed decline in active elders the two methods evaluated (Ridge L and NN-SLASSO by AMNR) showed that this pattern of abnormalities is characterized by a gradual and focused slowing of brain electrical activity in the supplementary, premotor and prefrontal motor regions, leading to increase of low frequency activity. The increase in the energy of the generators of theta activity in the supplementary and prefrontal areas in elders with slower gait speed, as compared with those with higher gait speed, could indicate the brain mechanisms reflecting the alterations in mobility in older adults. These mechanisms have been related both to the deficit in multisensory co-activation in these areas, which function as compensatory mechanisms for peripheral sensory deficit (Hawkins (2018)) and to the cognitive component of gait in older adults (Smith 2017). In our study, the NN-SLASSO presented sparser solutions with an easier interpretation of the sources involved in these mechanisms. Therefore, future studies should be carried out to evaluate the usefulness of using EEG source imaging correlates of these physical performance patterns to provide early indications of cognitive decline or adverse outcome in elders.

## 7. Conclusions

In this work we have introduced a modified Newton-Raphson (MNR) algorithm to estimate multiple penalized least squares (MPLS) models, and its extension to perform efficient optimization over the active set of selected features (AMNR). The proposed MNR technique can be interpreted as a generalization of the Majorize-Minimize (MM) algorithm to include combinations of constraints. The AMNR technique is an algorithm to estimate MPLS models in an active set framework using the direction of the MNR solutions as the direction descent, allowing for efficient implementation and offering strict sparse solutions. It also allows to naturally include sign constraints in addition to sparsity and/or smoothness, which leads to the introduction of new methods such as the Smooth NonNegative Garrote and NonNegative Smooth LASSO. We showed the usefulness of these new algorithms with simulation studies, especially their advantages to cope with highly underdetermined problems. We presented a preliminary exploration of its validity to estimate solutions to the EEG inverse problem using simulated and real experimental EEG data, finding that solutions obtained with the AMNR algorithm outperformed those with classical MNR techniques such as MM and LQA. Moreover, the new methods based on nonnegativity constraints showed promising results toward the improvement of localization and estimation of more focal sources. However, a full exploration of the validity of these methods to reliably localizing EEG sources in research and clinical applications is still needed. An interesting problem deserving future research is the development of AMNR algorithms to handle nonnegative solutions with methods such as Fusion and Fused LASSO or ENET. We would also like to explore the robustness to noise and doing more general assessments of the methods in the context of EEG/MEG source imaging by using other evaluation measures and by applying them to more complex scenarios.

**Supplementary Material**

The online supplementary material contains the derivation of the algorithms introduced in this work, as well as additional complementary details on the simulations and real data results.

## Acknowledgments

This work was supported by the VLIR-UOS project "A Cuban National School of Neurotechnology for Cognitive Aging" and the National Fund for Science and Innovation of Cuba. CU2017TEA436A103.

The Clinical Hospital of Chengdu Brain Science Institute, MOE Key Lab for Neuroinformation, University of Electronic Science and Technology of China, Chengdu, China; Cuban Center for Neurosciences, La Habana, Cuba.

E-mail: mayrim@cneuro.edu.cu

School of Computing and Intelligent Systems, Ulster University, UK.

E-mail: bornot@gmail.com

Cuban Center for Neurosciences, La Habana, Cuba.

E-mail: agustin@cneuro.edu.cu

Cuban Center for Neurosciences, La Habana, Cuba.

E-mail: jhoanna@cneuro.edu.cu

Cuban Center for Neurosciences, La Habana, Cuba.

E-mail: dariopalmeroledon@gmail.com

Cuban Center for Neurosciences, La Habana, Cuba.

E-mail: jose.alvarez@cneuro.edu.cu

Cuban Centre for Longevity, Ageing and Health Studies, Cuba.

E-mail: daysiga@infomed.sld.cu

Cuban Center for Neurosciences, La Habana, Cuba.

Ave. 25, esq. 158, Cubanacán, Havana City, Cuba

E-mail: eduardo@cneuro.edu.cu

The Clinical Hospital of Chengdu Brain Science Institute, MOE Key Lab for Neuroinformation, University of Electronic Science and Technology of China, Chengdu, China; Cuban Center for Neurosciences, La Habana, Cuba.

E-mail: pedro.valdes@neuroinformatics-collaboratory@cneuro.edu.cu

SUPPLEMENTARY MATERIAL

# Multiple penalized least squares and sign constraints with modified Newton-Raphson algorithms: application to EEG source imaging.

**Short running title**: MNR algorithms for EEG source imaging.

Mayrim Vega-Hernández[1,2,a], José M. Sánchez-Bornot[3,a], Agustín Lage Castellano[2], Jhoanna Pérez-Hidalgo-Gato[2], Darío Palmero-Ledón[2], José E. Alvarez Iglesias[2], Daysi García-Agustín[4], Eduardo Martínez-Montes[2,b], Pedro A. Valdés-Sosa[1,2,b].

[1] The Clinical Hospital of Chengdu Brain Science Institute, MOE Key Lab for Neuroinformation, University of Electronic Science and Technology of China, Chengdu, China.

[2] Cuban Center for Neurosciences, Havana, Cuba.

[3] School of Computing and Intelligent Systems, Ulster University, UK.

[4] Cuban Centre for Longevity, Ageing and Health Studies, Cuba.

[a] contributed equally to this research

[b] corresponding author.

## 1. Formulation of the Modified Newton-Raphson (MNR) algorithm and relation with the Majorize-Minimize (MM) algorithm

The Multiple Penalized Least Squares (MPLS) model (Vega-Hernández et al. (2008); Sánchez-Bornot et al. (2008)) is stated as follows:

$$\widehat{\boldsymbol{\beta}} = \underset{\boldsymbol{\beta}}{\mathrm{argmin}} \{f(\boldsymbol{\beta})\} = \underset{\boldsymbol{\beta}}{\mathrm{argmin}} \{ (\mathbf{y} - \mathbf{X}\boldsymbol{\beta})^T(\mathbf{y} - \mathbf{X}\boldsymbol{\beta}) + \Psi(\boldsymbol{\beta}) \} \quad (S1.1)$$

where the penalty term takes the form of a sum of several constraints or penalty functions, i.e., $\Psi(\boldsymbol{\beta}) = \sum_{r=1}^{R} \lambda_r \sum_{i=1}^{N_r} g_r\left(\left|\theta_i^{(r)}\right|\right)$. This is evaluated at the components of the vector $\boldsymbol{\theta}^{(r)} = \mathbf{L}^{(r)}\boldsymbol{\beta}$, with $\mathbf{L}^{(r)} \in \mathbb{R}^{N_r \times p}$ being linear operators that impose a structural relationship among coefficients, (e.g., the matrix of first or second differences). The regularization parameters $\lambda_r$, for $r = 1, \dots, R$, establish the relative importance of each constraint. As can be easily shown, LASSO and Ridge regression are instances of equation (S1.1) setting $R = 1$, $\mathbf{L} = \mathbf{I}_p$ (the $p \times p$ identity matrix) and using the L1 and L2 norms as penalty functions, respectively. These and other known particular examples of this general model are summarized in Table S1.

| Name | Penalty term | Function definition |
|---|---|---|
| Ridge I | $\Psi = \lambda \sum_{i=1}^{p} g(|\theta_i|)$ | $g(\theta) = \theta^2; \boldsymbol{\theta} = \boldsymbol{\beta}$ |
| Ridge L |  | $g(\theta) = \theta^2; \boldsymbol{\theta} = \mathbf{L}\boldsymbol{\beta}$ |
| LASSO |  | $g(\theta) = |\theta|; \boldsymbol{\theta} = \boldsymbol{\beta}$ |
| Fusion LASSO |  | $g(\theta) = |\theta|; \boldsymbol{\theta} = \mathbf{L}\boldsymbol{\beta}$ |
| Smooth LASSO (SLASSO) | $\Psi = \lambda_1 \sum_{i=1}^{p} g_1\left(\left|\theta_i^{(1)}\right|\right) + \lambda_2 \sum_{i=1}^{p} g_2\left(\left|\theta_i^{(2)}\right|\right)$ | $g_1(\theta) = |\theta|; \boldsymbol{\theta}^{(1)} = \boldsymbol{\beta}$ <br> $g_2(\theta) = \theta^2; \boldsymbol{\theta}^{(2)} = \boldsymbol{\Omega}\boldsymbol{\beta}$ |
| Elastic Net (ENET L) |  | $g_1(\theta) = |\theta|; \boldsymbol{\theta}^{(1)} = \mathbf{L}\boldsymbol{\beta}$ <br> $g_2(\theta) = \theta^2; \boldsymbol{\theta}^{(2)} = \mathbf{L}\boldsymbol{\beta}$ |
| Adaptive LASSO (ALASSO) | $\Psi = \lambda \sum_{i=1}^{p} \gamma_i g(|\theta_i|)$ | $g(\theta) = |\theta|; \boldsymbol{\theta} = \boldsymbol{\beta}$ <br> with $\gamma_i \geq 0$ for $i = 1, \ldots, p$ |
| Nonnegative Garrote (NNG) |  | $g(\theta) = |\theta|; \boldsymbol{\theta} = \boldsymbol{\beta}$ <br> with $\gamma_i = 1/\left|\beta_i^{ols}\right|$ <br> for $i = 1, \ldots, p$ |

**Table S1**: Known models represented as instances of the general MPLS model (equation S1.1). Here, $\boldsymbol{\Omega} \in \mathbb{R}^{p \times p}$ is the first-differences operator and $\mathbf{L} \in \mathbb{R}^{p \times p}$ is a matrix used for imposing a correlation structure in the solution, typically being the first- or second-differences operator. The $\boldsymbol{\beta}^{ols}$ is the ordinary least squares solution.

In order to derive a general modified Newton-Raphson algorithm for the MPLS model, it is important to find the gradient and the Hessian of the objective function for equation (S1.1):

$$\nabla f(\boldsymbol{\beta}) = -\mathbf{X}^T(\mathbf{y} - \mathbf{X}\boldsymbol{\beta}) + \sum_{r=1}^{R} \lambda_r \sum_{i=1}^{N_r} \nabla \theta_i^{(r)}(\boldsymbol{\beta}) g_r'\left(\left|\theta_i^{(r)}\right|\right) sgn\left(\theta_i^{(r)}\right)$$

$$= -\mathbf{X}^T(\mathbf{y} - \mathbf{X}\boldsymbol{\beta}) + \sum_{r=1}^{R} \lambda_r \sum_{i=1}^{N_r} \mathbf{L}_i^{(r)T}\left(g_r'\left(\left|\theta_i^{(r)}\right|\right) / \left|\theta_i^{(r)}\right|\right) \mathbf{L}_i^{(r)} \boldsymbol{\beta}$$

$$\nabla^2 f(\boldsymbol{\beta}) = \mathbf{X}^T\mathbf{X} + \sum_{r=1}^{R} \lambda_r \sum_{i=1}^{N_r} \mathbf{L}_i^{(r)T}\left(g_r'\left(\left|\theta_i^{(r)}\right|\right) / \left|\theta_i^{(r)}\right|\right) \mathbf{L}_i^{(r)}$$

where the scalar magnitude $\theta_i^{(r)}$ is the $i$-th element of $\boldsymbol{\theta}^{(r)}$ and $sgn(\cdot)$ represents the sign function, which was conveniently written as $sgn(x) = x/|x|$. $\mathbf{L}_i^{(r)}$ represents the $i$-th row of the matrix $\mathbf{L}^{(r)}$.

We now follow the same rationale used by the Majorize-Minimize (MM) algorithm of Hunter and Li (Hunter and Li (2005)). They showed that the local quadratic approximation is an instance of an MM algorithm. Therefore, we need to verify that our general penalty function $\Psi(\boldsymbol{\beta}) = \sum_{r=1}^{R} \lambda_r \sum_{i=1}^{N_r} g_r\left(\left|\theta_i^{(r)}\right|\right)$ satisfies the conditions established in Proposition 3.1 presented in (Hunter and Li (2005)) that we rewrite here with our notation for an easier understanding of its application to the general MPLS case.

**Proposition S1: (adapted from Hunter and Li, (2005)).** Suppose that on $(0, \infty)$, $g(\cdot)$ is piecewise differentiable, nondecreasing and convex. Furthermore, it is continuous at 0 and $g'(0_+) < \infty$ (i.e., the limit of $g'(x)$ as $x \to 0$ from positive values is finite). Then, for all $\theta_0 \neq 0$, the magnitude defined as: $\Phi_{\theta_0}(\theta) = g(|\theta_0|) + (\theta^2 - \theta_0^2)g'(|\theta_0|_+)/2|\theta_0|$, majorizes $g(|\theta|)$ and these two conditions hold: a) $\Phi_{\theta_0}(\theta) \geq g(|\theta|)$ for all $\theta$, with equality when $\theta = \pm|\theta_0|$; b) $\Phi_{\theta_0}(\theta) < \Phi_{\theta_0}(\theta_0)$ implies that $g(|\theta|) < g(|\theta_0|)$

Carrying out simple mathematical transformations, the penalty term of the objective function for MPLS models can be written as the sum of new functions:

$$\Psi(\boldsymbol{\beta}) = \sum_{r=1}^{R} \lambda_r \sum_{i=1}^{N_r} g_r\left(\left|\theta_i^{(r)}\right|\right) = \sum_{i=1}^{N_r} \sum_{r=1}^{R} \lambda_r g_r\left(\left|\theta_i^{(r)}\right|\right) = \sum_{i=1}^{N_r} \mathcal{G}(|\theta_i|)$$

In simple terms, this holds if all linear operators $\mathbf{L}^{(r)}$ are the same, such that the vector functions $\boldsymbol{\theta}^{(r)}(\boldsymbol{\beta}) = \mathbf{L}^{(r)}\boldsymbol{\beta}$ lead to the same variable with components $\theta_i^{(r)}$ for all $r$. In that case, it is easy to see that if every penalty function $g_r\left(\left|\theta_i^{(r)}\right|\right)$ satisfies the conditions of Proposition S1, then $\mathcal{G}(|\theta_i|) = \sum_{r=1}^{R} \lambda_r g_r\left(\left|\theta_i^{(r)}\right|\right)$ is piecewise differentiable on $(0, +\infty)$ and the results of Proposition 1 will hold. However, an algorithm for a more general case when the $\mathbf{L}^{(r)}$ are not the same, can be derived with the use of the local quadratic approximation proposed by Hunter and Li (Hunter and Li (2005)) for all functions $g_r$, which converts the original penalty term in an equivalent quadratic penalty. To avoid numerical problems when $\theta_i \approx 0$, they proposed to modify the objective function (i.e. an approximation $f_\varepsilon(\boldsymbol{\beta})$) by perturbing the penalty function $g$, using some small $\varepsilon > 0$, as: $g_\varepsilon(|\theta|) = g(|\theta|) - \varepsilon \int_0^{|\theta|} \frac{g'(t)}{\varepsilon + t} dt$.

Applying this perturbation to every penalty function $g_r$ in the MPLS model, we can obtain a local quadratic approximation for all of them as a function of a common variable $\theta$:

$$g_\varepsilon^{(r)}(|\theta|) \approx g_r\left(\left|\theta_i^{(r)}\right|\right) + \frac{\left(\theta^2 - \theta_i^{(r)2}\right) g_r'\left(\left|\theta_i^{(r)}\right|_+\right)}{2\left(\varepsilon + \left|\theta_i^{(r)}\right|\right)},$$

where the symbol $f(\theta_+)$ denotes the limit of $f(x)$ as $x \to \theta$ from positive values. The Newton-Raphson technique is then used to minimize the perturbed objective function through its first and second derivatives:

$$\nabla f_\varepsilon(\boldsymbol{\beta}) = -\mathbf{X}^T\mathbf{y} + \left(\mathbf{X}^T\mathbf{X} + \sum_{r=1}^{R} \lambda_r \mathbf{L}^{(r)T}\mathbf{D}^{(r)}\mathbf{L}^{(r)}\right)\boldsymbol{\beta}$$

$$\nabla^2 f_\varepsilon(\boldsymbol{\beta}) = \mathbf{X}^T\mathbf{X} + \sum_{r=1}^{R} \lambda_r \mathbf{L}^{(r)T}\mathbf{D}^{(r)}\mathbf{L}^{(r)}$$

where $\mathbf{D}^{(r)}$ is a diagonal matrix with diagonal elements defined as $d_i^{(r)} = g_r'\left(\left|\theta_i^{(r)}\right|\right)/\left(\varepsilon + \left|\theta_i^{(r)}\right|\right)$ for $i = 1, \ldots, N_r$ and some very small $\varepsilon > 0$. Then, we can locally minimize the perturbed objective function $f_\varepsilon(\boldsymbol{\beta})$ for some $\alpha_k > 0$, using the iterative formula:

$$\boldsymbol{\beta}_{k+1} = \boldsymbol{\beta}_k - \alpha_k\{\nabla^2 f_\varepsilon(\boldsymbol{\beta}_k)\}^{-1}\nabla f_\varepsilon(\boldsymbol{\beta}_k) = \boldsymbol{\beta}_k + \alpha_k\left[\left(\mathbf{X}^T\mathbf{X} + \sum_{r=1}^{R} \lambda_r \mathbf{L}^{(r)T}\mathbf{D}^{(r)}\mathbf{L}^{(r)}\right)^{-1}\mathbf{X}^T\mathbf{y} - \boldsymbol{\beta}_k\right]$$

Note that $\|f(\boldsymbol{\beta}) - f_\varepsilon(\boldsymbol{\beta})\| \to 0$ and $\|\nabla f(\boldsymbol{\beta}) - \nabla f_\varepsilon(\boldsymbol{\beta})\| \to 0$ uniformly whenever $\varepsilon \to 0$. Thus, any limit point of the estimated sequence $\boldsymbol{\beta}_1, \boldsymbol{\beta}_2, \ldots$ represent a critical point of the original objective function $f(\boldsymbol{\beta})$ (Hunter and Li, 2005).

In our case, the objective function may be more complex if it combines convex and concave penalty functions with correlation structure. In those cases, the NR algorithm can be stuck at saddle or local stationary points. However, the function $f(\boldsymbol{\beta})$ is convex for penalty functions based on L1 and L2 norms, e.g., Fused LASSO (Tibshirani et al. (2005)), Fusion LASSO (Land and Friedman (1996)) and Smooth LASSO (Hebiri and van de Geer (2011)). Therefore, for these cases the MNR implementation and, in particular, the canonical version ($\alpha_k = 1$), achieves the global minimum. Finally, the parameter $\varepsilon$ can be selected as proposed by Hunter and Li (Hunter and Li, 2005):

$$\varepsilon = \frac{tol}{2RM}\min\left\{\left|\theta_i^{(r)}\right| : \theta_i^{(r)} \neq 0\right\}, \text{ for } i = 1, \ldots, N_r \text{ and } r = 1, \ldots, R,$$

where $M = \max\{g_r'(0_+)\}$, for $r = 1, \ldots, R$, and $tol > 0$ is the convergence parameter (i.e. convergence is determined when an absolute change in every element of the vector solution is below a predefined value $tol$, such that $|\partial_j f_\varepsilon(\boldsymbol{\beta})| < tol/2$). The parameter $\varepsilon$ becomes smaller through iterations but it is usually fixed after the first iteration to avoid numerical instability (see Hansen (1998)).

## 2. Derivation of the Active-set Modified Newton-Raphson algorithm

Consider the MPLS optimization problem defined in its unconstrained variant:

$$\text{minimize } f(\boldsymbol{\beta}); \text{ with } f(\boldsymbol{\beta}) = \|\mathbf{y} - \mathbf{X}\boldsymbol{\beta}\|_2^2 + \Psi(\boldsymbol{\beta})$$

where $\Psi(\boldsymbol{\beta})$ is a sum of convex functions, which guarantees the convexity of the objective or cost function $f(\boldsymbol{\beta})$. An algorithm that solves this minimization problem can be a sequence of steps in the space of $\boldsymbol{\beta}$ that reduces the values of $f(\boldsymbol{\beta})$ until no more reduction can be achieved. Alternatively, the same algorithm can be interpreted as a sequence of steps in the space of $\boldsymbol{\beta}$ that solves the equation $\nabla f(\boldsymbol{\beta}) = 0$, as this become a sufficient condition of the global minimum when $f$ is a convex function. Let's assume we are at step $k$ of the minimization of $f(\boldsymbol{\beta})$ with coefficients vector $\boldsymbol{\beta}_k$ that will be updated as $\boldsymbol{\beta}_{k+1} = \boldsymbol{\beta}_k + \mathbf{b}_k$. If we take into account that the effect of previous steps can be absorbed by the residuals: $\mathbf{r}_k = \mathbf{y} - \mathbf{X}\boldsymbol{\beta}_k$, the cost function at step $k+1$ can be expressed as a function of the vector update $\mathbf{b}_k$:

$$f_{k+1}(\mathbf{b}_k) = \|\mathbf{y} - \mathbf{X}\boldsymbol{\beta}_{k+1}\|_2^2 + \Psi(\boldsymbol{\beta}_k + \mathbf{b}_k)$$

$$f_{k+1}(\mathbf{b}_k) = \|\mathbf{r}_k - \mathbf{X}\mathbf{b}_k\|_2^2 + \Psi(\boldsymbol{\beta}_k + \mathbf{b}_k)$$

The change in the cost function from the previous iteration $\left(f_k = f(\boldsymbol{\beta}_k) = f_{k+1}(\mathbf{b}_k)|_{\mathbf{b}_k = 0}\right)$ to the next iteration $(f_{k+1}(\mathbf{b}_k) = f(\boldsymbol{\beta}_k + \mathbf{b}_k))$, can be found from the second term in the first-order Taylor approximation of $f_{k+1}(\mathbf{b}_k)$ around $\mathbf{b}_k = 0$:

$$f_{k+1}(\mathbf{b}_k) \approx f_{k+1}(\mathbf{b}_k)|_{\mathbf{b}_k=0} + \left(\nabla_\mathbf{b} f_{k+1}(\mathbf{b}_k)|_{\mathbf{b}_k=0}\right)^T \mathbf{b}_k$$

$$f_{k+1}(\mathbf{b}_k) \approx f_k - \left(2\mathbf{x}^T \mathbf{r}_k - \nabla\Psi(\boldsymbol{\beta}_k)\right)^T \mathbf{b}_k \qquad (S2.1)$$

where we have used that the gradient of the continuous and differentiable penalty function with respect to $\mathbf{b}_k$, evaluated at $\mathbf{b}_k = 0$, is equal to the gradient with respect to $\boldsymbol{\beta}_k$, i.e. $\nabla_\mathbf{b}\Psi(\boldsymbol{\beta}_k + \mathbf{b}_k)|_{\mathbf{b}_k=0} = \nabla_{\boldsymbol{\beta}_k}\Psi(\boldsymbol{\beta}_k) = \nabla\Psi(\boldsymbol{\beta}_k)$, which is easy to verify.

Interestingly, the change in the cost function $f_{k+1} - f_k = -\left(2\mathbf{X}^T\mathbf{r}_k - \nabla\Psi(\boldsymbol{\beta}_k)\right)^T \mathbf{b}_k = -\nabla f(\boldsymbol{\beta}_k)^T(\boldsymbol{\beta}_{k+1} - \boldsymbol{\beta}_k)$, for small update vectors, is proportional to the product between the residuals and the predictors (columns of **X**) minus the gradient of the penalization term, and convergence will be achieved when these terms become equal. From this equation, we can then follow a procedure similar to that used to derive the LARS algorithm (Efron (2004)), in order to find the update (size and direction) $\mathbf{b}_k$ that ensures the minimization of the cost function in each iteration (i.e., $f_{k+1} \leq f_k$ for all $k$). This implies that $f_{k+1} - f_k \leq 0$, which leads to the local-global minimum condition $\nabla f(\boldsymbol{\beta}_k)^T(\boldsymbol{\beta}_{k+1} - \boldsymbol{\beta}_k) \geq 0$. In the case that $\boldsymbol{\beta}_k$ reaches a minimum of the cost function, say $\boldsymbol{\beta}_k = \boldsymbol{\beta}^*$, this is a sufficient condition for $\boldsymbol{\beta}^*$ to be a global minimum of $f$ over the convex set $S$, where the cost function is continuously differentiable and its gradient is continuous (Bertsekas, (1997)). From this condition, and a proper definition of $S$, a set of three optimality conditions are obtained, i.e., a set of sufficient conditions for the estimator $\boldsymbol{\beta}_k$ at each iteration to be a local minimum, which means that such algorithm will provide the path of optimal solutions along iterations. The formal introduction and demonstration of a Proposition presenting the local-global minimum condition and the three optimality conditions for the Adaptive Lasso model are presented in Section 3 of this Supplementary Material (Proposition S2 and Theorem S1).

The optimality conditions allow us to derive a general Active-set Modified Newton-Raphson (AMNR) algorithm for MPLS models, as they impose a relationship between the gradient of the cost function and its minimum. For instance, the first optimality condition implies that nonzero coefficients (components of the vector solution included in the active set) and their corresponding derivative of the cost function (components of the gradient), must have opposite signs. The other two optimality conditions imply that the absolute value of the gradient for those coefficients in the active set is the same and is the highest among all predictors. These conditions can be achieved by selecting in every iteration one coefficient to be included in the active set, as the one with the highest absolute value of the gradient, while using other empirical procedures to ensure that the sign of the gradient is opposite to that of the coefficient itself. In particular, a general LARS-type constraint to be fulfilled in every iteration $k$ is:

$$|\nabla f(\boldsymbol{\beta}_k)| = |2\mathbf{X}_\mathcal{A}^T \mathbf{r}_k - \nabla \Psi(\boldsymbol{\beta}_k)| = C_{max} \mathbf{1}_\mathcal{A} \text{ with } C_{max} > 0$$

where the active set is represented as $\mathcal{A}$, the columns of the corresponding predictors form the matrix $\mathbf{X}_\mathcal{A}$, and $\mathbf{1}_\mathcal{A}$ represents a vector of ones with length equal to the cardinality of the active set (later denoted by $|\mathcal{A}|$). This implies that the absolute value of the gradient for all coefficients included in the active set will be equal to the value $C_{max}$. The common use of the absolute value in this condition in LARS aims at controlling the sign of the updated coefficients and ensuring that the last term in equation (S2.1) is positive (Efron (2004)). We propose that a more natural way of taking this into account is by explicitly including the sign of the solution in the equation (i.e., $sgn(\nabla f(\boldsymbol{\beta}_k)) = -sgn(\boldsymbol{\beta}_k)$, as imposed by the first optimality condition):

$$\nabla f(\boldsymbol{\beta}_k) = |\nabla f(\boldsymbol{\beta}_k)| sgn(\nabla f(\boldsymbol{\beta}_k))$$

$$-\left(2\mathbf{X}_{\mathcal{A}}^T \mathbf{r}_k - \nabla \Psi(\boldsymbol{\beta}_k)\right) = -C_{max} sgn(\boldsymbol{\beta}_k)$$

$$2\mathbf{X}_{\mathcal{A}}^T \mathbf{r}_k - \nabla \Psi(\boldsymbol{\beta}_k) = C_{max} sgn(\boldsymbol{\beta}_k) \quad (S2.2)$$

where we define the component-wise sign function of a vector ($sgn(\mathbf{x})$) as returning a vector containing the sign of each component of the argument ($\mathbf{x}$).

The derivation of the more general MPLS problem deserves a subsequent theoretical paper. One approach is to make use of the quadratic approximation of the MNR procedure explained in the previous Section. Briefly, an MPLS model consisting of multiple penalties as proposed in Section 1 of this Supplementary Material $\left(\Psi(\boldsymbol{\beta}) = \sum_{r=1}^{R} \lambda_r \sum_{i=1}^{N_r} g_r\left(\left|\theta_i^{(r)}\right|\right)\right)$, can be rewritten as a general quadratic model by using the quadratic approximations of $g_r$ functions; i.e. obtaining $\Psi(\boldsymbol{\beta}) = \|\mathbf{W}\boldsymbol{\beta}\|_2^2$ (where $\mathbf{W} = \left(\sum_{r=1}^{R} \lambda_r \mathbf{L}^{(r)T} \mathbf{D}^{(r)} \mathbf{L}^{(r)}\right)^{1/2}$ is a general matrix that combines all linear operators for the different penalty functions). However, this also means that any model formed by a combination of penalty functions based on L1 and L2 norms can be taken to a simple LASSO or Adaptive LASSO model, by joining all other penalties (except one based on the L1-norm) into a quadratic term and using the trick of data augmentation (Hebiri and van de Geer (2011)). Therefore, in this work we derive and implement one of the simplest versions of the AMNR algorithm, which corresponds to the Adaptive LASSO model, conveniently established on the constrained equivalent formulation:

$$\text{minimize} \|\mathbf{y} - \mathbf{X}\boldsymbol{\beta}\|_2^2 \text{ subject to } \sum_{j=1}^{p} \gamma_j |\beta_j| \leq \tau,$$

where $\gamma_j$ are positive weights, thus reducing this model to LASSO when they are all set to 1. This formulation avoids the explicit use of regularization parameters by replacing them by a thresholding parameter ($\tau > 0$).

For this model, the AMNR implementation is very similar to the LARS algorithm (see Efron et al. 2004) for LASSO; however, it does not require predictors to be standardized and can also be used to minimize a continuously differentiable objective function while imposing sign constraints over the parameters. As mentioned above, the algorithm is an active-set procedure, which means that -at every iteration- only a subset of predictors is included in the active set and updated, while the others remain at zero. In this particular case, as in LARS, the active set is updated at every iteration in one of two ways: 1) including a new predictor or 2) excluding an existent one; such that the active sets in two consecutive iterations differ only by one predictor. If the optimality conditions are satisfied within the algorithm, the solutions will always comply with the constraint of the optimization problem ($\sum_{j=1}^{p} \gamma_j |\beta_j| \leq \tau$). Therefore, in practice, we do not need to check that this condition holds and it is usually easier to use a simpler stop criterion based on the convergence of the absolute value of the gradient to a value smaller than a tolerance ($tol$). Although it is not straightforward, there exist a relationship such that we can fix a value for $\tau$ and find the corresponding tolerance to be used in the algorithm (i.e., hard constraint). If the $tol$ is fixed first, then the level of the constraint $\tau$ will depend on the values attained by coefficients in the final solution (i.e., soft constraint).

We show next that the AMNR technique for this model guarantees that optimality conditions in Theorem S1 (Section 3), are sufficient conditions for the estimator at each iteration to be a local minimum, and thus allows obtaining the path of optimal solutions.

Writing equation S2.2 for the Adaptive LASSO model:

$$\mathbf{X}_{\mathcal{A}}^T \mathbf{r}_k = C_k \mathbf{\Gamma}_{\mathcal{A}} sgn(\boldsymbol{\beta}_k) \tag{S2.3}$$

where $\mathbf{\Gamma}_{\mathcal{A}}$ is the diagonal matrix of factors $\{\gamma_j : j \in \mathcal{A}\}$ and $C_{max} = 2C_k \mathbf{\Gamma}_{\mathcal{A}}$ is represented through $C_k = \max(|\mathbf{\Gamma}^{-1} \mathbf{X}^T \mathbf{r}_k|)$ in this iteration, which ensures fulfilling the second and third optimality conditions (b and c in Section 3 of this Supplementary Material). At this point we write the update vector as $\mathbf{b}_k = \alpha_k \boldsymbol{\delta}_k$ to explicitly separate it in size ($\alpha_k$) and direction ($\boldsymbol{\delta}_k$), with the latter being nonzero for those coefficients in the active set, thus we will work with $\boldsymbol{\delta}_k$ and $\boldsymbol{\beta}_k$ as vectors with only $|\mathcal{A}|$ components in each iteration. Then, we write equation S2.3 for iteration $k+1$, and substitute residuals $\mathbf{r}_{k+1} = \mathbf{y} - \mathbf{X}_{\mathcal{A}}\boldsymbol{\beta}_{k+1} = \mathbf{y} - \mathbf{X}_{\mathcal{A}}\boldsymbol{\beta}_k - \mathbf{X}_{\mathcal{A}}\mathbf{b}_k = \mathbf{r}_k - \alpha_k \mathbf{X}_{\mathcal{A}}\boldsymbol{\delta}_k$, for those coefficients included in the active set:

$$\mathbf{X}_{\mathcal{A}}^T \mathbf{r}_k - \alpha_k \mathbf{X}_{\mathcal{A}}^T \mathbf{X}_{\mathcal{A}} \boldsymbol{\delta}_k = C_{k+1} \boldsymbol{\Gamma}_{\mathcal{A}} sgn(\boldsymbol{\beta}_{k+1})$$

$$C_k \boldsymbol{\Gamma}_{\mathcal{A}} sgn(\boldsymbol{\beta}_k) - \alpha_k \mathbf{X}_{\mathcal{A}}^T \mathbf{X}_{\mathcal{A}} \boldsymbol{\delta}_k = C_{k+1} \boldsymbol{\Gamma}_{\mathcal{A}} sgn(\boldsymbol{\beta}_{k+1})$$

$$\alpha_k \mathbf{X}_{\mathcal{A}}^T \mathbf{X}_{\mathcal{A}} \boldsymbol{\delta}_k = C_k \boldsymbol{\Gamma}_{\mathcal{A}} sgn(\boldsymbol{\beta}_k) - C_{k+1} \boldsymbol{\Gamma}_{\mathcal{A}} sgn(\boldsymbol{\beta}_{k+1})$$

$$\alpha_k \mathbf{X}_{\mathcal{A}}^T \mathbf{X}_{\mathcal{A}} \boldsymbol{\delta}_k = (C_k - C_{k+1}) \boldsymbol{\Gamma}_{\mathcal{A}} sgn(\boldsymbol{\beta}_k)$$

The last step in previous derivation assumed that the sign of coefficients included in the active set do not change from one iteration to the next one. This can be ensured within the algorithm, as will be explained later on. From this equation we can solve for both the size and direction by splitting it in vectorial and scalar part. The vectorial part leads to the Newton-Raphson direction when conveniently dividing by $C_k$ in both sides, and using S2.3 again:

$$\boldsymbol{\delta}_k = (\mathbf{X}_{\mathcal{A}}^T \mathbf{X}_{\mathcal{A}})^{-1} C_k \boldsymbol{\Gamma}_{\mathcal{A}} sgn(\boldsymbol{\beta}_k)$$

$$\boldsymbol{\delta}_k = (\mathbf{X}_{\mathcal{A}}^T \mathbf{X}_{\mathcal{A}})^{-1} \mathbf{X}_{\mathcal{A}}^T \mathbf{r}_k$$

while the scalar part leads to $\alpha_k = (C_k - C_{k+1})/C_k$, or equivalently $C_{k+1} = C_k(1 - \alpha_k)$.

This means that our algorithm will use the Newton-Raphson direction to update the coefficients in the active set and move over the space of optimal solutions $\boldsymbol{\beta}_{k+1} = \boldsymbol{\beta}_k + \alpha \boldsymbol{\delta}_k$ for some $\alpha \in (0, \alpha_k]$, where $0 < \alpha_k \leq 1$ and $\boldsymbol{\beta}_{k+1}$ not including here any new predictor yet. As noted above, this will comply with the optimality conditions only if we ensure that the signs of all coefficients in the active set do not change ($sgn(\boldsymbol{\beta}_k) = sign(\boldsymbol{\beta}_{k+1})$). Therefore, at every iteration we must check if exists a step size $\alpha \in (0, \alpha_k]$ such that any active coefficient becomes zero, i.e. if for any coefficient $i \in \mathcal{A}$, it holds that $\beta_{k_i} + \alpha \delta_{k_i} = 0$. This leads to compute, at iteration $k$, the value $\alpha_k^0 = \min^+ \left\{ \alpha_j^0 = -\beta_{k_j}/\delta_{k_j} : \beta_{k_j}(\beta_{k_j} + \delta_{k_j}) < 0, j \in \mathcal{A} \right\}$, where $\min^+$ indicates that the minimum is taken considering only the positive values for computed elements.

For those non-active predictors ($\mathbf{x}_j : j \in \mathcal{A}^C$), equation S2.3 is not valid at iteration $k$, but it must be valid for the new predictor selected to join the active set at iteration $k+1$. Therefore, we must select $\mathbf{x}_j$ such that $\mathbf{x}_j^T \mathbf{r}_k - \alpha_k \mathbf{x}_j^T \mathbf{X}_{\mathcal{A}} \boldsymbol{\delta}_k = C_{k+1} \gamma_j sgn\left(\beta_{k+1_j}\right)$, which also ensures that the selected coefficient $\beta_{k+1_j}$ and the corresponding correlation between predictor and residual ($-\mathbf{x}_j^T \mathbf{r}_{k+1}$) have opposite signs, in accordance with the first optimality condition. From this condition, using $C_{k+1} = C_k(1-\alpha_k)$ and a shorter notation for the correlation of the predictor with residuals ($c_{k_j} = \mathbf{x}_j^T \mathbf{r}_k$) and with the projection of the update direction ($a_{k_j} = \mathbf{x}_j^T \mathbf{X}_{\mathcal{A}} \boldsymbol{\delta}_k$), we can obtain an analytical expression for the step size $\alpha_k$ needed for computing the coefficient corresponding to the non-active predictor $\beta_{k+1_j}$ to be included in the active set:

$$c_{k_j} - \alpha_k a_{k_j} = C_{k+1} \gamma_j sgn(\beta_{k+1_j})$$

$$c_{k_j} - \alpha_k a_{k_j} = C_k(1-\alpha_k)\gamma_j sgn(\beta_{k+1_j})$$

$$\alpha_k = \frac{C_k \gamma_j sgn(\beta_{k+1_j}) - c_{k_j}}{C_k \gamma_j sgn(\beta_{k+1_j}) - a_{k_j}}$$

The sign of the coefficient to be computed is not known a priori, but this expression implies that the step sizes needed to enter a coefficient in the active set with positive or negative values will be different:

$$\alpha_k^+ = \frac{C_k \gamma_j - c_{k_j}}{C_k \gamma_j - a_{k_j}}$$

$$\alpha_k^- = \frac{-C_k \gamma_j - c_{k_j}}{-C_k \gamma_j - a_{k_j}} = \frac{C_k \gamma_j + c_{k_j}}{C_k \gamma_j + a_{k_j}}$$

Therefore, as long as step sizes are valid, i.e., they are positive, this means that if -in any iteration *k*- the minimum step size for a predictor to join the active set corresponded to $\alpha_k^+$ ($\alpha_k^-$), then the corresponding coefficient will take a positive (negative) value when included in the active set. This is a very useful result, as it allows us to have a natural way to impose nonnegative or nonpositive constraints. For instance, if we want to find the best positive (negative) solution we can just ignore the step sizes that will lead to negative (positive) coefficients and include in the active set only coefficients that will enter with a positive (negative) value. As we are also ensuring that all coefficients in the active set will keep the same sign or removed from the active set if they tend to change sign, it turns out that all coefficients in the final solution will have the same sign and we will get properly sign-constrained solutions.

Finally, we recall that the fundamental property of the algorithm is that the gradient of the objective function $|2\mathbf{X}_{\mathcal{A}}^T \mathbf{r}_k|$ monotonically decreases along the selected step direction with a

positive step size. As a result, there will always be a step size ensuring that a new predictor, not included in the active set, joins the active set. This step size should be taken to be the minimum among all predictors, no matter if it is for including a negative or a positive coefficient. However, another important point is to keep the signs of the coefficients already in the active set unchanged, therefore, the step size for which one of these coefficients becomes zero $\alpha_k^0$ should be used if it is smaller than $\alpha_k^+$ and $\alpha_k^-$. With these considerations, the final estimate for the step size becomes

$$\alpha_k = \min{}^+ \left\{ \alpha_k^0, \alpha_{k\,j}^+, \alpha_{k\,j}^- : j \in \mathcal{A}_k^C \right\}.$$

Following these considerations and the discussion above, an AMNR algorithm for the Adaptive LASSO model can be summarized as:

---

**AMNR algorithm for Adaptive LASSO** ($\mathbf{y} \in \mathbb{R}^{n \times 1}, \mathbf{X} \in \mathbb{R}^{n \times p}, \gamma_1, \ldots, \gamma_p$)

---

1. Initialization

    Start with $k \leftarrow 0, tol \leftarrow 10^{-8}, \mathcal{A} \leftarrow \{\ \}, Z \leftarrow \{1, 2, \ldots, p\}$, and $\boldsymbol{\beta}_0 \leftarrow \mathbf{0}_p$.

2. Set $k \leftarrow k + 1$.

    Compute $\mathbf{r}_k \leftarrow \mathbf{y} - \mathbf{X}^T \boldsymbol{\beta}_{k-1}$ and $\mathbf{c} \leftarrow \mathbf{X}^T \mathbf{r}_k$.

3. If stop condition is true, **go to Step 9**

    If the set $Z$ is empty or if $|c_j|/\gamma_j \leq tol$ for all $j \in Z$, then **go to Step 9.**

4. Select the index of a new predictor $j_k \in Z$ to be included in the active set $\mathcal{A}$

    Find an index $j_k \in Z$ such that $|c_{j_k}|/\gamma_{j_k} = \max\{|c_j|/\gamma_j : j \in Z\}$. Set $C_k \leftarrow |c_{j_k}|/\gamma_{j_k}$.

5. Move $j_k$ from $Z$ to $\mathcal{A}$, the insertion order must be conserved, that is $\mathcal{A} \leftarrow [\mathcal{A}, \{j_k\}]$ and $Z \leftarrow Z \setminus \{j_k\}$

6. Let $\mathbf{X}_{\mathcal{A}}$ denote the matrix of active predictors.

    Compute $\boldsymbol{\delta} \leftarrow (\mathbf{X}_{\mathcal{A}}^T \mathbf{X}_{\mathcal{A}})^{-1} \mathbf{X}_{\mathcal{A}}^T \mathbf{r}_k$

7. Compute the step $\alpha$ in the direction of the gradient descent $\boldsymbol{\delta}$

    a) Compute $\mathbf{a} \leftarrow \mathbf{X}^T \mathbf{X}_{\mathcal{A}} \boldsymbol{\delta}$.

    b) Find an index $h \in Z$ such that

    $$(C_h \gamma_h - c_h)/(C_h \gamma_h - a_h) = \min\{(C_k \gamma_j - c_j)/(C_k \gamma_j - a_j) : a_j < C_k \gamma_j, j \in Z\}.$$

    If $h$ exist, then set $\alpha^+ \leftarrow (C_h \gamma_h - c_h)/(C_h \gamma_h - a_h)$, else set $\alpha^+ \leftarrow 1$.

    c) Find an index $h \in Z$ such that

$$(C_h\gamma_h + c_h)/(C_h\gamma_h + a_h) = \min\{(C_k\gamma_j + c_j)/(C_k\gamma_j + a_j): a_j > -C_k\gamma_j, j \in Z\}.$$

If $h$ exist, then set $\alpha^- \leftarrow (C_h\gamma_h + c_h)/(C_h\gamma_h + a_h)$, else set $\alpha^- \leftarrow 1$.

d) Find an index $j_h \in \mathcal{A}$ such that

$$-\beta_{k-1 j_h}/\delta_h = \min\{-\beta_{k-1 j_l}/\delta_l : \beta_{k-1 j_l}(\beta_{k-1 j_l} + \delta_l) < 0, j_l \in \mathcal{A}\}.$$

If $j_h$ exist, then set $\alpha^0 \leftarrow -\beta_{k-1 j_h}/\delta_h$, else set $\alpha^0 \leftarrow 1$.

e) Set $\alpha \leftarrow \min\{\alpha^+, \alpha^-, \alpha^0\}$.

8. Compute $\beta_{kj} \leftarrow \beta_{k-1 j} + \alpha \delta_j$ for all $j \in \mathcal{A}$.

   a) If $\alpha = 1$ or $\alpha < \alpha^0$, then **go to Step 2**.

   b) Move from set $\mathcal{A}$ to set $Z$ all indices $j_l \in \mathcal{A}$ for which $\beta_{k j_l} = 0$.

   c) Set $k \leftarrow k + 1$.

   d) Compute $\mathbf{r}_k \leftarrow \mathbf{y} - \mathbf{X}^T \boldsymbol{\beta}_{k-1}$ and $\mathbf{c} \leftarrow \mathbf{X}^T \mathbf{r}_k$. Find an index $h \in Z$ such that

   $$|c_h|/\gamma_h = \max\{|c_j|/\gamma_j : j \in Z\}. \text{ Set } C_k \leftarrow |c_h|/\gamma_h. \text{ **Go to** Step 6}.$$

9. Ending step: the whole path of optimal solutions $\boldsymbol{\beta}_0, \boldsymbol{\beta}_1, \ldots, \boldsymbol{\beta}_k$ is computed.

---

## 3. Optimality conditions for Adaptive Lasso

**Proposition S2: (local-global minimum conditions)**

Let $f(\boldsymbol{\beta}) : \mathbb{R}^{p \times 1} \to \mathbb{R}$ be a continuously differentiable function defined on the convex set $S \subseteq \mathbb{R}^p$, with continuous derivative $\nabla f : \mathbb{R}^{p \times 1} \to \mathbb{R}^{p \times 1}$

a) If $\boldsymbol{\beta}^*$ is a local minimum of $f$ over $S \subset \mathbb{R}^p$, then

$$\nabla f(\boldsymbol{\beta}^*)^T (\boldsymbol{\beta} - \boldsymbol{\beta}^*) \geq 0, \qquad \forall \boldsymbol{\beta} \in S$$

b) If $f(\boldsymbol{\beta})$ is convex over $S$, then the condition a) is a sufficient condition for $\boldsymbol{\beta}^*$ to be a global minimum of $f$ over $S$.

**Proof:** See (Bertsekas, (1997))

**Theorem S1 (necessary conditions for optimal estimator for Adaptive LASSO):**

Consider the optimization problem $\widehat{\boldsymbol{\beta}} = \underset{\boldsymbol{\beta}}{\operatorname{argmin}}\{f(\boldsymbol{\beta})\}$, subject to $\sum \gamma_j |\beta_j| \leq \tau$; where $f(\boldsymbol{\beta})$ is continuously differentiable, all weights are positive $\gamma_j > 0$, $\forall j$, and $\tau > 0$ is a given scalar. If $\boldsymbol{\beta}^*$ satisfies the conditions of Proposition S2, then the following properties hold:

a) If $\beta_j^* > 0$ then $\partial_j f(\boldsymbol{\beta}^*) \leq 0$, and if $\beta_j^* < 0$ then $\partial_j f(\boldsymbol{\beta}^*) \geq 0$.

b) If $\beta_i^* = 0$ and $\beta_j^* \neq 0$ then $\gamma_j |\partial_i f(\boldsymbol{\beta}^*)| \leq \gamma_i |\partial_j f(\boldsymbol{\beta}^*)|$.

c) If $\beta_i^* \neq 0$ and $\beta_j^* \neq 0$ for $i \neq j$, then $\gamma_j |\partial_i f(\boldsymbol{\beta}^*)| = \gamma_i |\partial_j f(\boldsymbol{\beta}^*)|$.

**Proof:** According to the conditions of Proposition 1 for this problem, the optimal solution $\boldsymbol{\beta}^*$ satisfies:

$$\sum_j \partial_i f(\boldsymbol{\beta}^*)(\beta_j - \beta_j^*) \geq 0, \ \forall \boldsymbol{\beta} \in S$$

In order to verify a), suppose that $\beta_j^* > 0$ for some $j = 1, \ldots, p$ and $\boldsymbol{\beta}$ is a feasible solution to the problem (that is $\sum \gamma_j |\beta_j| \leq t$). For example, take $\beta_i = \beta_i^*$, for each $i \neq j$, and $\beta_j = \beta_j^* - \varepsilon$, for $0 < \varepsilon < \beta_j^*$. Therefore, if we apply proposition 1, we get inequality $-\varepsilon \partial_j f(\boldsymbol{\beta}^*) \geq 0$, which is equivalent to $\partial_j f(\boldsymbol{\beta}^*) \leq 0$. Similarly, $\beta_j^* < 0 \Rightarrow \partial_j f(\boldsymbol{\beta}^*) \geq 0$ is demonstrated.

To prove b), suppose that $\beta_i^* = 0$ and $\beta_j^* \neq 0$, in particular $\beta_j^* > 0$. Take a feasible solution such that $\beta_i = \varepsilon/\gamma_i$, $\beta_j = \beta_j^* - \varepsilon/\gamma_j$ and $\beta_k = \beta_k^*$ for each $k \neq i, j$ and some $0 < \varepsilon < \gamma_j \beta_j^*$. Applying the conditions in Proposition 1, we obtain $\varepsilon(\partial_i f(\boldsymbol{\beta}^*)/\gamma_i - \partial_j f(\boldsymbol{\beta}^*)/\gamma_j) \geq 0$, which implies that $\gamma_j \partial_i f(\boldsymbol{\beta}^*) \geq \gamma_i \partial_j f(\boldsymbol{\beta}^*)$. Similarly, by taking $\beta_i = -\varepsilon/\gamma_i$, we obtain $-\varepsilon(\partial_i f(\boldsymbol{\beta}^*)/\gamma_i - \partial_j f(\boldsymbol{\beta}^*)/\gamma_j) \geq 0$, which implies that $-\gamma_j \partial_i f(\boldsymbol{\beta}^*) \geq \gamma_i \partial_j f(\boldsymbol{\beta}^*)$. From these we conclude that $\gamma_i \partial_j f(\boldsymbol{\beta}^*) \leq -\gamma_j |\partial_i f(\boldsymbol{\beta}^*)|$. On the other hand, if $\beta_j^* < 0$ is chosen with the same considerations, then we can demonstrate that $\gamma_i \partial_j f(\boldsymbol{\beta}^*) \geq \gamma_j |\partial_i f(\boldsymbol{\beta}^*)|$. These conclusions together imply that $\gamma_i |\partial_j f(\boldsymbol{\beta}^*)| \geq \gamma_j |\partial_i f(\boldsymbol{\beta}^*)|$ and thus, condition b) is demonstrated.

In order to demonstrate c), suppose that $\beta_i^* > 0$ and $\beta_j^* > 0$ for $i \neq j$. Also suppose that a feasible solution $\boldsymbol{\beta}$ exist such that $\beta_i = \beta_i^* + \varepsilon/\gamma_i$ and $\beta_j = \beta_j^* - \varepsilon/\gamma_j$, for some $0 < \varepsilon < \gamma_j \beta_j^*$, and $\beta_k = \beta_k^*$, for each $k \neq i, j$. Then, if we apply the first condition of Proposition 1, we obtain $\gamma_j \partial_i f(\boldsymbol{\beta}^*) \geq \gamma_i \partial_j f(\boldsymbol{\beta}^*)$. This inequality is symmetric with respect to the selected indices; therefore, all positive components at the optimal solution have minimal (negative) and equal partial cost derivatives. With similar reasoning but selecting $\beta_i^* < 0$ and $\beta_j^* < 0$ instead, we conclude that all negative components have maximal (positive) and equal partial cost derivatives. Finally, in the situation that $\beta_i^* < 0$ and $\beta_j^* > 0$ for $i \neq j$, take $\boldsymbol{\beta} \in S$ such that that $\beta_i = \beta_i^* - \varepsilon/\gamma_i$ and $\beta_j = \beta_j^* - \varepsilon/\gamma_j$, for any $0 < \varepsilon < \gamma_j \beta_j^*$ and $\beta_k = \beta_k^*$ for each $k \neq i, j$. By using Proposition 1, we obtain that $-\varepsilon(\partial_i f(\boldsymbol{\beta}^*)/\gamma_i + \partial_j f(\boldsymbol{\beta}^*)/\gamma_j) \geq 0$, which implies that $-\gamma_j \partial_i f(\boldsymbol{\beta}^*) \geq \gamma_i \partial_j f(\boldsymbol{\beta}^*)$. Taking $\beta_i = \beta_i^* + \varepsilon/\gamma_i$ and $\beta_j = \beta_j^* + \varepsilon/\gamma_j$, for some $0 < \varepsilon < \gamma_i \beta_i^*$ and $\beta_k = \beta_k^*$ for each $k \neq i, j$, in the same conditions, we obtain that $\varepsilon(\partial_i f(\boldsymbol{\beta}^*)/\gamma_i + \partial_j f(\boldsymbol{\beta}^*)/\gamma_j) \geq 0$, which implies that $\gamma_i \partial_j f(\boldsymbol{\beta}^*) \geq -\gamma_j \partial_i f(\boldsymbol{\beta}^*)$. Therefore $\gamma_i |\partial_j f(\boldsymbol{\beta}^*)| = \gamma_j |\partial_i f(\boldsymbol{\beta}^*)|$ and thus, condition c) is satisfied. ∎

## 4. Simulation study

Figure S1 presents a boxplot of the computational time (in seconds), showing that -besides the non-iterative Ridge solutions- the faster models are LASSO using LARS algorithm and NN-SLASSO and NNG using AMNR for whatever reference estimator. As expected, the same models using AMNR were generally faster than when computed with the MNR algorithm.

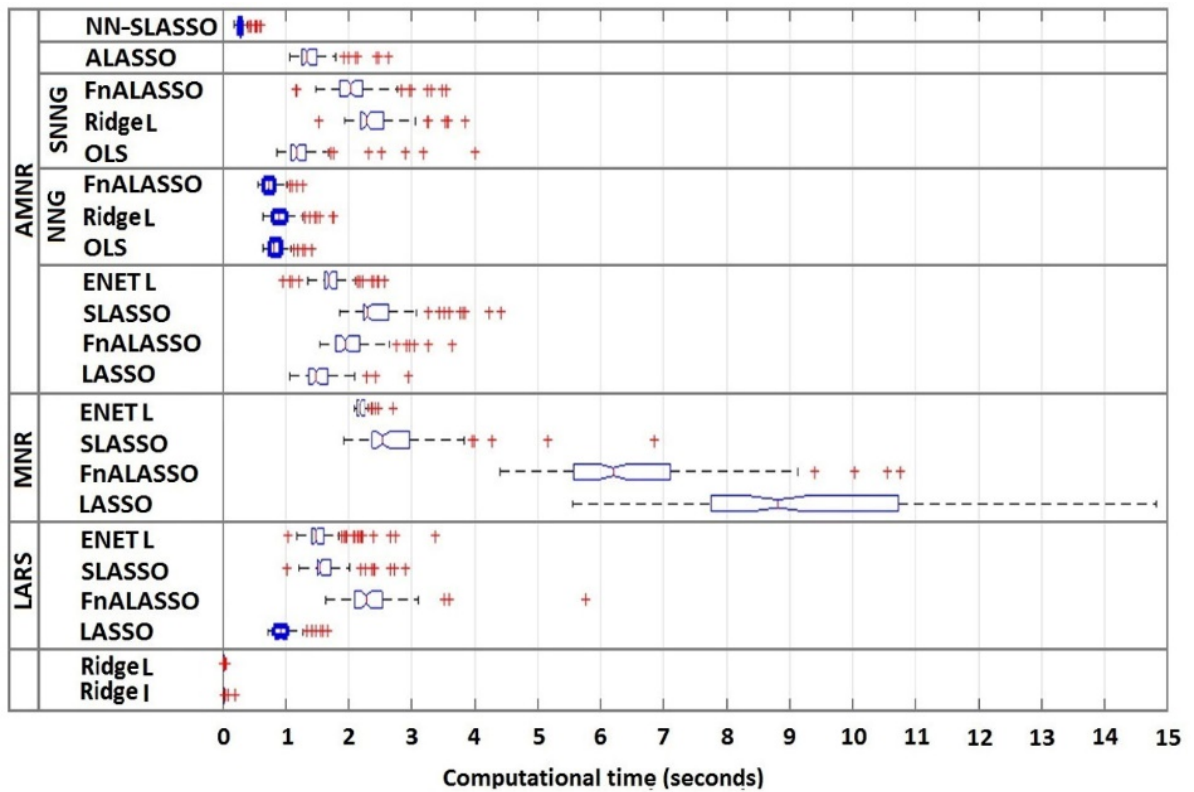

**Figure S1** Boxplots of the time necessary for computing one solution in each combination of model and algorithm, from the simulations using $p = 200$ and $n = 100$. The algorithm for estimating Ridge solutions is not mentioned as it is a simple evaluation of the regularized inverse (Tikhonov) (Tikhonov et al. (1995)).

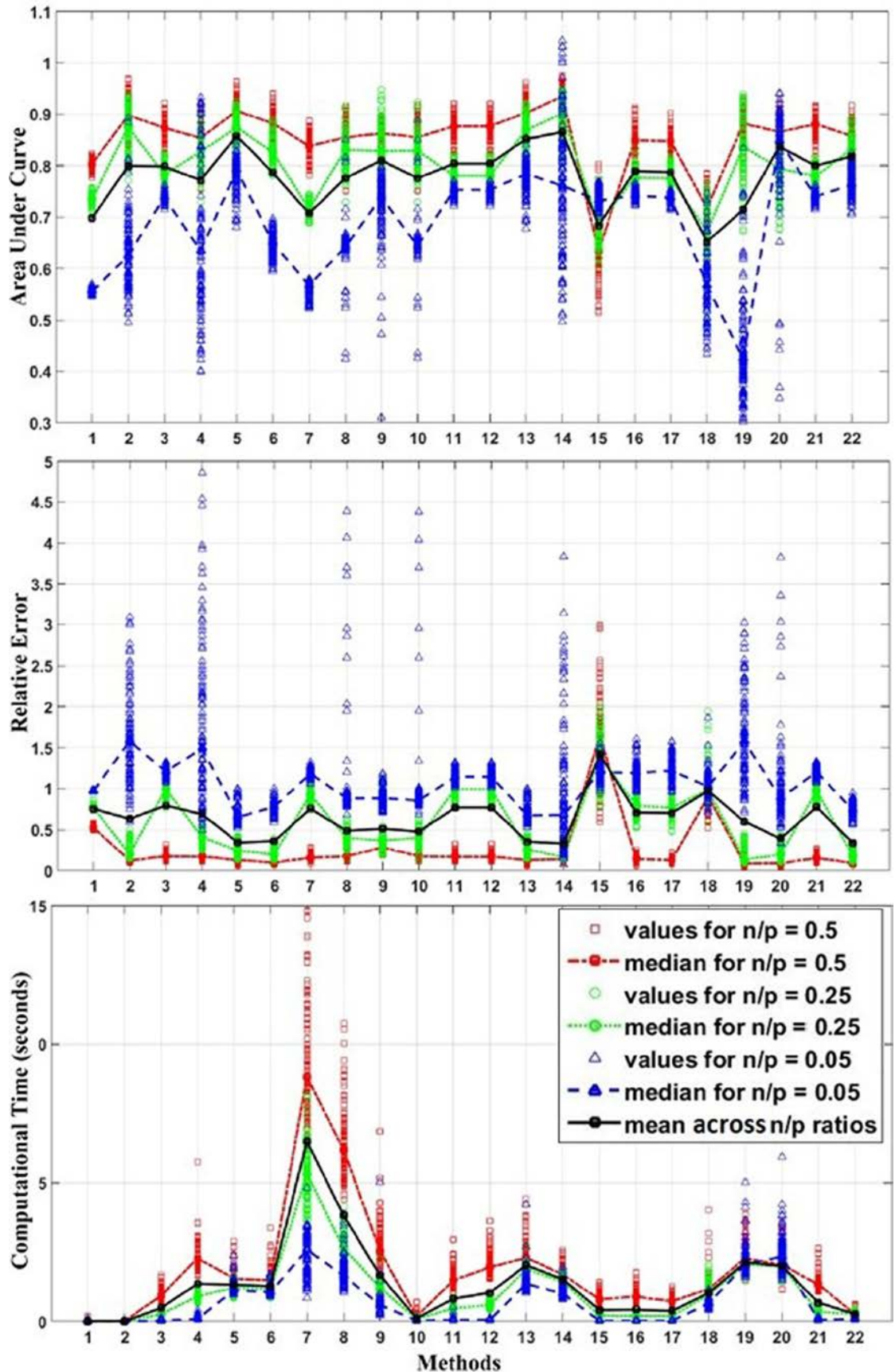

**Figure S2:** Medians of evaluation criteria (Relative Error, AUC and computational time) obtained from all 100 estimated solutions for the three different n/p ratios. The black line represents the mean value across all n/p ratios. The 22 methods correspond to combination of models and algorithms as numbered in Table 2 in the main text.

Figure S2 shows the median of AUC, median Relative Error and median of the computation time across the 100 estimated solutions in the three cases of n/p ratios {0.05, 0.25, 0.5} and the mean across these three n/p ratios (black line). The methods with better behavior for any n/p relation were SLASSO (with LARS and AMNR algorithms), ENET L (with LARS and AMNR algorithms), SNNG (with FnLASSO as reference estimator) and the nonnegative version of SLASSO (NN-SLASSO). These results are consistent with the qualitative pictures given in Figure 1 and Figure 2 (both in the main text), where SLASSO, ENET L, SNNG and NN-SLASSO are the solutions that better reconstructed the simulated bell and square regions. It is clear that the general difficulties in reconstructing the point source will not be largely reflected in the quantitative measures, as it is just one out of 200 estimated points. The analysis of the time necessary for computing one solution showed that the fastest methods are Ridge I and Ridge L, NNG (with the three reference solutions) and NN-SLASSO.

## 5. EEG Simulated data

The synthetic data consisted in four different sets of simulated primary current density (PCD) distributions, all of them simulated as a three-dimensional Gaussian source with amplitude of 10 nA/mm$^2$ and width of 10 mm (spherical). Each set contain seven PCDs: a 'centroid' PCD with maximum located in a particular anatomical structure of a brain space of 3862 generators, and 6 others derived from this one by locating the maxima in each of the 6 closest neighbor generators. Maximum values of the simulated PCDs were located in 1) the cingulate region left (Cingulate), 2) occipital pole left (Occipital), 3) postcentral gyrus (Postcentral), and temporal gyrus right (Temporal) as shown in the first row of Figure S3. Talairach Coordinates (Talairach and Tournoux (1988)) of the maximum value of each simulated PCD appear in Table S2.

Figure S3 shows the estimated sources by the best methods according to Table 2 in the main text, corresponding to the simulated 'centroid' PCDs in each region. We also added the Ridge L solution, which is mathematically equivalent to a classical solution known as LORETA in the field of EEG source localization (Pascual-Marqui et al (1994)). As expected, the Ridge L solutions are very smooth, while ENET L and SNNG methods (computed with AMNR) offered solutions that fluctuate between different degrees of sparsity/smoothness. Also, the use of sign constraints (allowed by AMNR) in the new inverse solutions SNNG and NN-SLASSO, led to sparser solutions than the unconstrained counterparts. SNNG solutions seem to be sparser versions of the reference solutions but without removing all ghost sources. The NN-SLASSO solutions are over-sparse but showing much less ghost sources as a convenient side effect. This solution also improves the localization of the main source with respect to ENET L, offering a very good localization even for the deepest simulated PCD (Temporal).

| | **Coordinate** | | | | **Coordinate** | | | | **Coordinate** | | | | **Coordinate** | | |
|---|---|---|---|---|---|---|---|---|---|---|---|---|---|---|---|
| | x | y | z | | x | y | z | | x | y | z | | x | y | Z |
| **Region: Cingulate** | **-8** | **48** | **5** | **Region: Postcentral** | **20** | **-43** | **68** | **Region: Occipital** | **-22** | **-99** | **-2** | **Region: Temporal** | **41** | **-8** | **-37** |
| | 6 | 48 | 5 | | 13 | -43 | 68 | | -29 | -99 | -2 | | 34 | -8 | -37 |
| | -1 | 48 | 5 | | 27 | -43 | 68 | | -15 | -99 | -2 | | 34 | 13 | -30 |
| | -8 | 48 | 12 | | 20 | -43 | 75 | | -22 | -99 | 5 | | 41 | -8 | -30 |
| | -8 | 48 | -2 | | 20 | -43 | 61 | | -22 | -99 | -9 | | 41 | -8 | -44 |
| | -8 | 55 | 5 | | 20 | -36 | 68 | | -22 | -92 | -2 | | 41 | -1 | -37 |
| | -8 | 41 | 5 | | 13 | -43 | 61 | | -22 | -92 | -9 | | 41 | -15 | -37 |

**Table S2:** Talairach coordinates (Talairach and Tournoux (1988)) of the maximum value of simulated solutions. The first row in each case shows the coordinates of the "centroid" simulated PCD (bold).

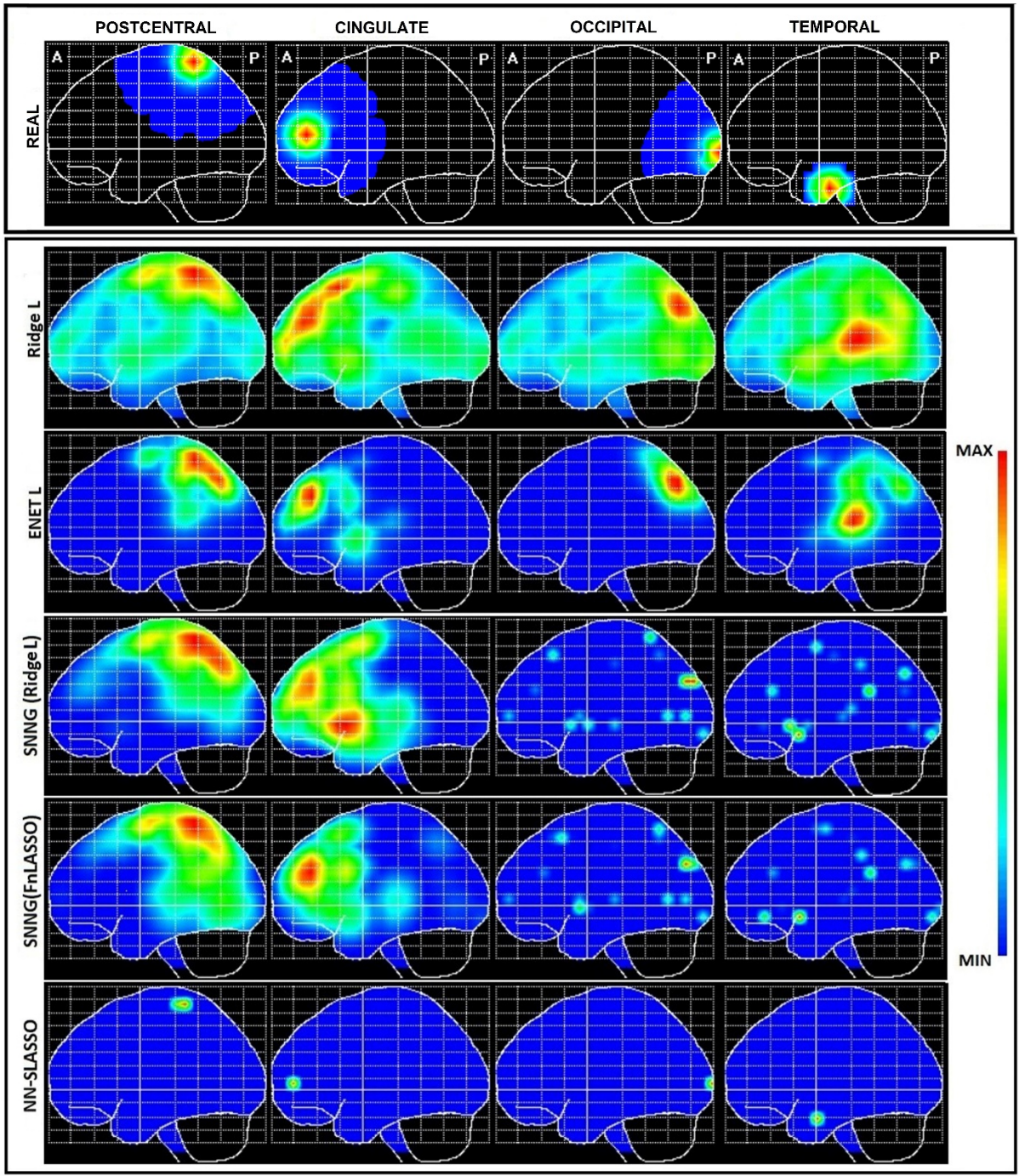

**Figure S3:** Maximum intensity projection in the sagittal plane of the four simulated 'centroid' PCDs (top row) and the corresponding estimated PCDs using five different methods (Ridge L; ENET L, SNNG with two different reference estimators and NN-SLASSO, the last four computed using the AMNR algorithm).